\documentclass[a4paper, aps, floatfix, footinbib, amsmath, amssymb, pra, twocolumn, superscriptaddress, longbibliography, showkeys]{revtex4-2}

\usepackage[english]{babel}
\usepackage[utf8]{inputenc}
\usepackage[colorinlistoftodos, color=green!40, prependcaption]{todonotes}
\usepackage{amsthm}
\usepackage{mathtools}
\usepackage{physics}
\usepackage{xcolor}
\usepackage{graphicx}
\usepackage[caption=false]{subfig}
\usepackage[left=23mm,right=13mm,top=35mm,columnsep=15pt]{geometry} 
\usepackage{adjustbox}
\usepackage{placeins}
\usepackage[T1]{fontenc}
\usepackage{lipsum}
\usepackage{csquotes}
\usepackage{amsmath,bm} % Math packages

\newtheorem{result}{Result}

\usepackage[pdftex, pdftitle={Article}, pdfauthor={Author}]{hyperref}
\usepackage[final]{microtype}

\begin{document}

\title{Preservability of Measurement Incompatibility: Purification, Activation, and a No-Go Theorem}

\author{Chao-Hsien Wu}
%\email{chaohsianwu@gmail.com}
\affiliation{Department of Physics, National Taiwan Normal University, Taipei 11677, Taiwan}
% Your name
    
%\author{Chung-Yun Hsieh}
%\affiliation{H. H. Wills Physics Laboratory, University of Bristol, Tyndall Avenue, Bristol, BS81TL, UK}

%\author{Benjamin Stratton}
%    \affiliation{Quantum Engineering Centre for Doctoral Training, H. H. Wills Physics Laboratory and Department of Electrical \& Electronic Engineering, University of Bristol, BS81FD, UK}

\author{Franco Nori}
\affiliation{Center of Quantum Computing, RIKEN, Wakoshi, Saitama, 351-0198, Japan}
\affiliation{Department of Physics, The University of Michigan, Ann Arbor, 48109-1040 Michigan, USA}

\author{Huan-Yu Ku}
    \email[]{huan.yu@ntnu.edu.tw}% Your name
    \affiliation{Department of Physics, National Taiwan Normal University, Taipei 11677, Taiwan}

\date{\today} % Leave empty to omit a date

\begin{abstract}
Measurement incompatibility is a fundamental quantum resource that enables advantages in many quantum information tasks, including cryptography and communication. However, unavoidable interactions between a system and its environment can degrade or even completely destroy measurement incompatibility; such a process is referred to as a measurement-incompatibility-annihilating channel. Recently, the capability of noisy quantum dynamics to preserve measurement incompatibility has been characterized within the resource theory of measurement incompatibility preservability.
This motivates studying how to purify the preservation of measurement incompatibility and how to activate it from a measurement-incompatibility-annihilating channel. To this end, we first introduce our figures of merit as robustness-based resource monotones within this resource theory. We demonstrate that while pre-filtering operations can strengthen measurement incompatibility preservability, they cannot activate it from an incompatibility-annihilating channel, establishing a no-go theorem. Furthermore, we explicitly demonstrate that an incompatibility-annihilating channel can be stochastically activated via post-filtering operations. Our results provide a practical framework for exploiting measurement incompatibility in quantum information processing.

\end{abstract}
%\keywords{first keyword, second keyword, third keyword}

\maketitle

\section{Introduction} \label{sec:introduction}
    One of the milestones in quantum mechanics is the measurement uncertainty principle, namely, the momentum and the position of a quantum system cannot be simultaneously measured without disturbance~\cite{BuschRMP}. In quantum information theory, this limitation is captured more generally by the notion of measurement incompatibility, which characterizes whether a set of measurements can be implemented jointly through a single measurement procedure~\cite{Heinosaari_2016,GuhneRMP2023}. Measurement incompatibility plays a crucial role in quantum information theory, as it constitutes a necessary resource for demonstrating quantum advantages, e.g., Bell nonlocality~\cite{BrunnerRMP,QuintinoPRL2014,ChenPRL2016,CavalcantiPRA2016,Ducuara2026PRL}, quantum steering~\cite{Hsieh2016PRA,Uola2020RMP,Zhao2020npj,Hsieh2023arXiv}, and quantum thermodynamics~\cite{HsiehPRL2024,Hsieh2024arXiv}.
    Given its operational significance, it is natural to ask how measurement incompatibility behaves under noisy dynamics.

    %Recently, the measurement measurement incompatibility preservability of a noisy channel has been studied. 
    %The characterization and quantification of a channel's measurement incompatibility preservability are proposed under the resource theory framework
    This motivates the question of whether, and to what extent, a noisy quantum channel can maintain measurement incompatibility. This ability, termed measurement incompatibility preservability, can be characterized and quantified within the framework of quantum resource theories~\cite{EricRMP2019,Hsieh2025PRA}. 
    In a general resource-theoretic setting, a resource-annihilating channel completely destroys the resource of interest. A representative example is an entanglement-annihilating channel, which maps every input state to a state without entanglement~\cite{Moravcikova_2010}. In contrast, a channel that retains the resource for at least some inputs can be regarded as resource-preserving~\cite{Hsieh2020resource,HsiehPRR2020,HsiehPRXQ2021,TakahashiPRA2022,StrattonPRL2024}.
    %\Note{Note that in general, a resource-annihilating channel is different from a resource-breaking channel (see concrete example for entanglement scenario~\cite{RossetPRX2018,ChenPRA2021,KuPRXQ2022,Lucas2025PRR}).}
    %\Note{[CH: Since we do not mention resource-breaking channels later in the article, I wonder if it is necessary to discuss the distinction between resource-annihilating and resource-breaking channels here.]}

    In the context of measurement incompatibility, an incompatibility-annihilating (IA) channel is one that maps every set of measurements to jointly measurable. In other words, a measurement-incompatibility-annihilating channel destroys measurement incompatibility for any set of measurements. In this work, we focus on IA channels and ask whether filtering operations can enable them to preserve measurement incompatibility. Such a phenomenon is known as resource activation~\cite{SanduPRL1995,HirschPRL2013,QuintinohysPRA2016,KuPRR2021,KuPRA2023,VillegasAguilar2024,Gelo2024arXiv}, where suitable quantum operations allow an initially useless object to exhibit a useful resource.
    
    Beyond activation, we also ask whether filtering can further enhance the measurement incompatibility preservability of a channel. We refer to this enhancement as purification of measurement incompatibility preservability.
    
    To address these questions, we develop a framework for quantifying and manipulating how well a quantum channel preserves measurement incompatibility. Specifically, we introduce a computable robustness measure for measurement incompatibility preservability and establish an operational formulation in terms of quantum steering. These results provide efficient tools for characterizing and optimizing the measurement incompatibility preservability of a channel.
    
    Using this framework, we establish a no-go theorem showing that pre-filtering operation alone can never activate an IA channel. We further show that post-filtering operations can stochastically activate IA channels, while the combination of pre- and post-filtering operations enables both activation and purification within a single protocol.
    
    %In this work, we study: (1) how an incompatibility-annihilating (IA) channel assisted with filtering operations can preserve measurement incompatibility. This phenomenon is called resource activation since the given useless resource can become useful via some quantum operations~\cite{SanduPRL1995,HirschPRL2013,QuintinohysPRA2016,KuPRR2021,KuPRA2023,VillegasAguilar2024,Gelo2024arXiv}. (2) How measurement incompatibility preservability can be further strengthened via a filtering operation. More specifically, we develop an operational framework for manipulating the measurement incompatibility preservability of quantum channels. We introduce a computable robustness monotone together with an equivalent steering-based formulation, allowing measurement incompatibility preservability to be efficiently characterized and optimized. This framework enables us to establish a no-go theorem proving that IA channels cannot be activated by pre-filtering operations alone. We further show that post-filtering operations can stochastically activate IA channels, while combining pre- and post-filtering operations realizes both activation and distillation within a single protocol.
    
%\section{Result}
%\label{sec:result}
\section{Measurement incompatibility and measurement incompatibility preservability}
    
    Measurement incompatibility is a fundamental feature of quantum mechanics, reflecting the essence of the uncertainty principle, i.e., one cannot measure physical quantities simultaneously, such as the position and momentum of a quantum particle. Specifically, a measurement is mathematically described by a set of positive-operator valued measurements (POVMs) satisfying $M_{a}\geq 0$ and $\sum_a M_a=\openone$. A set of measurements $\mathbf{M} \coloneqq \{M_{a|x}\}_{a,x}$, called a measurement assemblage, can then be characterized by a collection of POVMs, such that $\sum_{a}M_{a|x}=\openone~\forall~x$.
    
    A measurement assemblage is called jointly measurable or compatible if there exists a single parent POVM $\{G_\lambda\}_\lambda$ together with a classical post-processing map $p(a|x,\lambda)$ satisfying $M_{a|x}=\sum_\lambda p(a|x, \lambda)G_\lambda$ for all $a, x$. In this case, the measurement outcomes for different settings $x$ can be reproduced from the parent measurement ${G_\lambda}$ with purely classical data-processing. The set of compatible measurement assemblages is denoted as $\mathbb{JM}$. Conversely, if no such decomposition exists, the measurement assemblage is incompatible.
    %Measurement incompatibility can be quantified through incompatibility robustness~\cite{XXX}:
    %\begin{equation}
    %    \begin{aligned}
    %        &\mathcal{R}_I(\mathbf{M}) \coloneqq\\ 
    %        &\min_{t} \left\{ t \geq 0 \left| \frac{\left(M_{a|x}\right)+t\left(E_{a|x}\right)}{1+t} = M^{\text{JM}}_{a|x}, \quad \forall a, x \right\}\right.
    %        \label{eq:IR}
    %    \end{aligned}
    %\end{equation}
    %\Note{command out this}
    %where $\mathbf{E}\coloneqq\{E_{a|x}\}_{a, x}$ is arbitrary measurement assemblage and $\left\{M^{\text{JM}}_{a|x}\right\}_{a, x} \in \mathbb{JM}$. In general, robustness quantifies the minimal amount of noise required to destroy a given resource. In the case of incompatibility robustness, the resource under consideration is the incompatibility of a set of measurements.

\begin{figure}
         \centering
         \includegraphics[width=\linewidth]{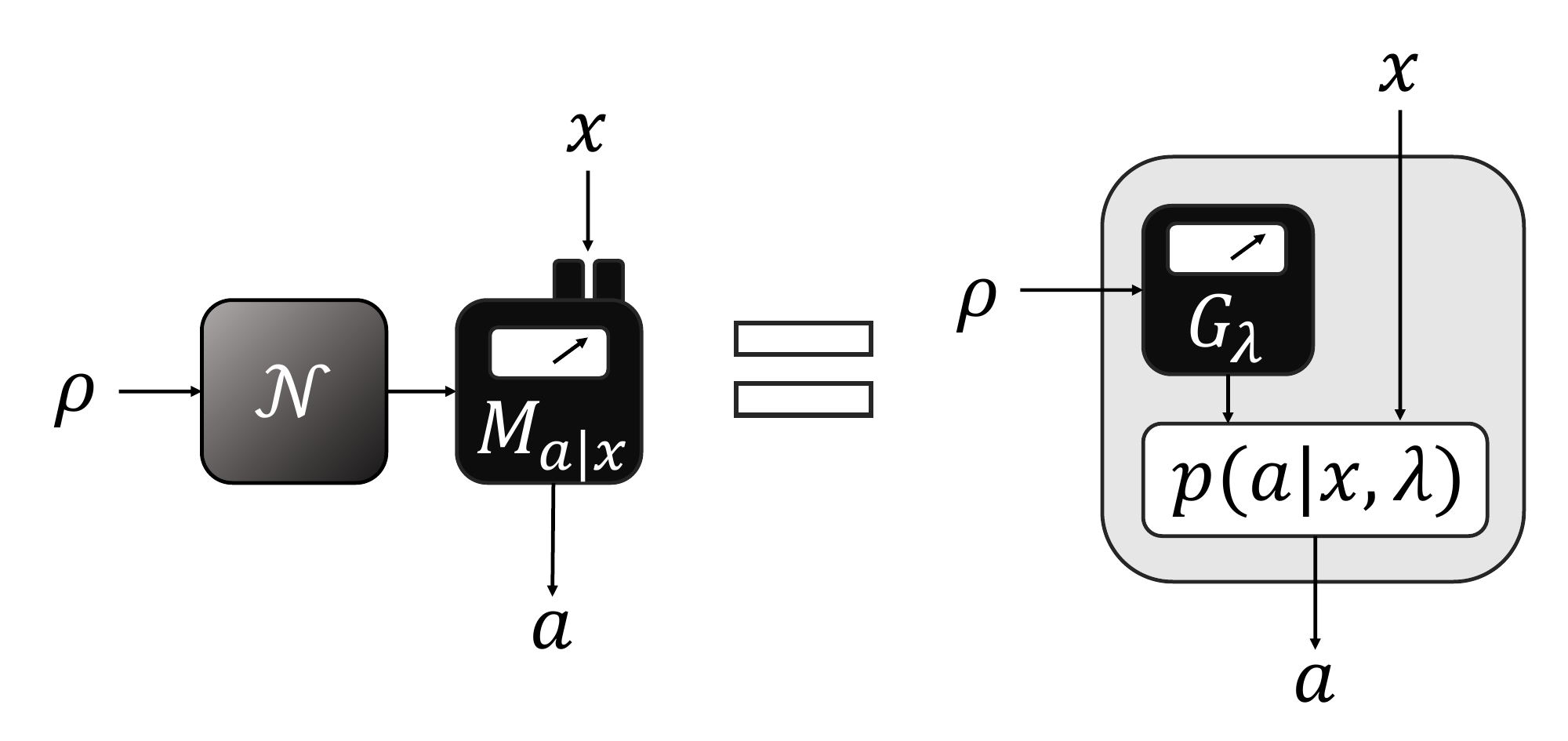}
         \caption{Scheme of IA channels. A channel \(\mathcal{N}\) is incompatibility annihilating if, after applying the channel, the measurement statistic \(\operatorname{Tr}\left[\rho\mathcal{N}^\dagger(M_{a|x})\right]\) can always reproduced by using a jointly measurable measurements \(\sum_\lambda p(a|x, \lambda)\operatorname{Tr}[\rho G_{\lambda}]\).} 
         \label{fig:IA channel}
     \end{figure}

    To study the ability of a measurement incompatibility within a noisy channel, one can use measurement incompatibility preservability to characterize the action of a quantum channel on a measurement assemblage. Specifically, a channel $\mathcal{N}$ is said to be $|\mathcal{X}|$-incompatibility-annihilating if, for every measurement assemblage $\{M_{a|x}\}_{a,x}$ with number of setting $|\mathcal{X}|$, there must exist a JM assemblage $M^{\text{JM}}_{a|x}$ such that $\mathcal{N}^{\dagger}(M_{a|x})=M^{\text{JM}}_{a|x}$ (see Fig.~\ref{fig:IA channel}). The set of IA channels is denoted as $\mathbb{IA}$. A channel possesses measurement incompatibility preservability if it is not IA.
    
    In recent work~\cite{Hsieh2025PRA}, a resource-theoretic framework for measurement incompatibility preservability is proposed, in which quantum channels are treated as resources when they preserve measurement incompatibility. Within this framework, the set of allowed operations, denoted as $\mathbb{AO}$, consists of supermaps that cannot transform an IA channel into a non-IA one, a.k.a. measurement incompatibility preservability.
    Specifically, an allowed operation $\Lambda$ must satisfy $\Lambda(\mathcal{N}) \in \mathbb{IA}$ for all $\mathcal{N} \in \mathbb{IA}$, meaning that any transformation that maps IA channels to IA channels is regarded as free. %\HY{[HY: Free operation needs to write down.]}

\section{Quantifying measurement incompatibility preserving channel}
    \label{subsec:quantifying}
    A robustness-based measure for measurement incompatibility preservability $\mathcal{R}_{\text{IP}}$ is defined as~\cite{Hsieh2025PRA}    
    \begin{eqnarray}
        \mathcal{R}_{\text{IP}}(\mathcal{N}) \coloneqq \min_{t, \mathcal{W}}\left\{t\geq0\left|\frac{\mathcal{N}+t\mathcal{W}}{1+t}\in\mathbb{IA}, \mathcal{W}\in \mathbb{C}\right\}\right.,
    \label{eq:robustness IA}
    \end{eqnarray}
    where $\mathbb{C}$ is the set of all quantum channels. Measurement incompatibility preservability robustness quantifies the smallest amount of noise strength $t$ such that the noisy mixture $\left(\mathcal{N}+t\mathcal{W}\right)/\left(1+t\right)$ of the channel completely destroys measurement incompatibility for any measurement assemblage. Therefore, $\mathcal{R}_{\text{IP}}(\mathcal{N})$ geometrically measures the minimal distance between the given channel $\mathcal{N}$ and the set $\mathbb{IA}$.
    
    However, since we cannot mathematically characterize the set of IA channels, Eq.~\eqref{eq:robustness IA} is not computable. To address this issue, we reformulate the definition of measurement incompatibility preservability robustness by testing whether the output measurement assemblage is always JM.
    By the definition of IA channels, the condition $
    \frac{\mathcal{N}+t\mathcal{W}}{1+t}\in\mathbb{IA}$ is equivalent to $\frac{\mathcal{N}^{\dagger}+t\mathcal{W}^{\dagger}}{1+t}\left(M_{a|x}\right) = M^{\text{JM}}_{a|x}$ for all $\mathbf{M}$, where $\left\{M^{\text{JM}}_{a|x}\right\}_{a, x} \in \mathbb{JM}$. Thus, for a fixed measurement assemblage $\mathbf{M}$, we can define the following optimization problem
    \begin{equation}
        \begin{aligned}
            &\mathcal{R}_{\text{IP}}^{\text{JM}}(\mathcal{N}; \mathbf{M}) \coloneqq\\ 
            &\min_{t, \mathcal{W}} \left\{t\geq0 \middle| \frac{\mathcal{N}^{\dagger}+t\mathcal{W}^{\dagger}}{1+t}\left(M_{a|x}\right) = M^{\text{JM}}_{a|x}, \quad \forall a, x \right\} .
            \label{eq:optimization JM}
        \end{aligned}
    \end{equation}
    By introducing the Choi–Jamiolkowski isomorphism of quantum channels: for an arbitrary channel \(\mathcal{C} \in \mathbb{C}\), we have
    \begin{equation}
        \begin{aligned}
            &\mathcal{C}(X) = \operatorname{Tr}_{\text{in}}[(\openone_{\text{out}} \otimes X^T)C] \\
            &\mathcal{C}^\dagger(X) = \operatorname{Tr}_{\text{out}}[(X^T \otimes \openone_{\text{in}})C]
            \label{eq:choi representation}
        \end{aligned}
    \end{equation}
    where \(C = \sum_{i, j} \mathcal{C}(\ket{i}\bra{j})_{\text{out}} \otimes \ket{i}\bra{j}_{\text{in}} \) is the corresponding Choi operator of channel \(\mathcal{C}\) and satisfies \(\operatorname{Tr}_{\text{out}}(C) = \openone_{\text{in}}\).
    
    The optimization problem in Eq.~\eqref{eq:optimization JM} can be solved efficiently via a semidefinite program (SDP) by defining $W$ as the Choi representation of $\mathcal{W}$ and $(1+t)G_\lambda$ as the parent measurement. We denote its primal and dual problems (see Appendix~\ref{appendix:dual problem derivation} for the derivation of dual problem):
    \begin{equation}
        \begin{aligned}
            \textbf{Primal: }
            \text{Given } &\mathcal{N}, \mathbf{M} \\
            \text{Find } &\min_{t, W, \{G_\lambda\}_{\lambda}} t = \mathcal{R}^{\text{JM}}_{\text{IP}}(\mathcal{N}; \mathbf{M})\\
            \text{s.t. } &\mathcal{N}^\dagger(M_{a|x}) + \operatorname{Tr}_{\text{out}}[(M_{a|x}^T \otimes \openone_{\text{in}})W] \\
            &= \sum_\lambda D(a|x, \lambda)G_\lambda, \quad \forall a, x \\
            &\operatorname{Tr}_{\text{out}}(W) = t\openone_{\text{in}}, \quad W \geq 0 \\
            &G_\lambda \geq 0, \quad \forall\lambda \\
            &\sum_\lambda G_\lambda = (1+t)\openone \\
            & t \geq 0
            \label{eq:optimization JM SDP}.
        \end{aligned}
    \end{equation}
    \begin{equation}
        \begin{aligned}
            \textbf{Dual: }
            \text{Given } &\mathcal{N}, \mathbf{M}\\
            \text{Find } &\max_{\mathbf{Y}, Z, S} \sum_{a, x}\operatorname{Tr}[Y_{a|x}\mathcal{N}^\dagger(M_{a|x})] - 1 \\
            &=\mathcal{R}^{\text{JM}}_{\text{IP}}(\mathcal{N}; \mathbf{M})\\
            \text{s.t. } & Y_{a|x} = Y^\dagger_{a|x}, \quad \forall a, x\\
            &\sum_{a, x} M^T_{a|x} \otimes Y_{a|x} \geq \openone \otimes S \\
            &\operatorname{Tr}[S] \geq 0, \quad S = S^\dagger\\
            &Z \geq \sum_{a, x} D(a|x, \lambda)Y_{a|x} \\
            &\operatorname{Tr}[Z] \leq 1, \quad Z = Z^\dagger
            \label{eq:optimization JM SDP dual}.
        \end{aligned}
    \end{equation}
    
    Here, \(D(a|x, \lambda)= \delta_{a, \lambda_x}\) is the deterministic response functions. By further optimizing over the input measurement assemblage, we obtain a new robustness of measurement incompatibility preservability:
    \begin{equation}
        \begin{aligned}
            \mathcal{R}_{\text{IP}}^{\text{JM}}(\mathcal{N}) \coloneqq \max_{\mathbf{M}} \mathcal{R}_{\text{IP}}^{\text{JM}}(\mathcal{N}; \mathbf{M}).
            \label{eq:robustness JM}
        \end{aligned}
    \end{equation}
    This quantity characterizes the maximal amount of measurement incompatibility that can be preserved by the channel $\mathcal{N}$, where the maximization is taken over all measurement assemblages. Importantly, this optimized robustness is not merely an operationally motivated quantity, but also satisfies the basic requirements of a resource monotone in the resource theory. This gives our first main result.
    \begin{result}
        ${\mathcal{R}_{\text{IP}}^{\text{JM}}}(\mathcal{N})$ is a valid resource monotone under the resource theory of measurement incompatibility preservability; e.g. it satisfies the following properties:
        \begin{eqnarray}
            \begin{aligned}
                \text{1. }&\textit{Faithfulness: } \mathcal{R}_{\text{IP}}^{\text{JM}}(\mathcal{N}) = 0 \iff \forall \mathcal{N} \in \mathbb{IA}. \\
                \text{2. }&\textit{Monotonicity: } \\ &\mathcal{R}_{\text{IP}}^{\text{JM}}(\Lambda(\mathcal{N}))\leq \mathcal{R}_{\text{IP}}^{\text{JM}}(\mathcal{N}), \forall \Lambda \in \mathbb{AO}. \\
                \text{3. }&\textit{Convexity: } \\ &\mathcal{R}_{\text{IP}}^{\text{JM}}\left( \sum_{i}p(i)\mathcal{N}_{i} \right) \leq \sum_{i}p(i)\mathcal{R}_{\text{IP}}^{\text{JM}}\left( \mathcal{N}_{i}\right).
                \label{eq:resource monotone}
            \end{aligned}
        \end{eqnarray}
        \label{result:monotone}
    \end{result}
    The proof of Result~\ref{result:monotone} is provided in Appendix~\ref{appendix:monotone proof}. Although now $\mathcal{R}_{\text{IP}}^{\text{JM}}$ is not an SDP due to the maximization of the measurement assemblage, we adopt the see-saw algorithm (see Appendix~\ref{appendix:seesaw}) to potentially find the optimal measurement assemblage in Eq.~\eqref{eq:robustness JM}.

\section{Measurement incompatibility preservability robustness in a steering scenario}

    A natural question is whether an IA channel can be "activated" into a measurement incompatibility preserving channel. In studies of activation phenomena, filters are often employed as an experimentally accessible device. A filter $\mathcal{F}$ is a probabilistic operation described by a completely positive (CP), trace–non-increasing (TNI) map.
    
    In the Schr\"odinger picture, the action of a filtering operation is well defined: $\rho \mapsto \mathcal{F}(\rho)/\operatorname{Tr}[(\mathcal{F}(\rho)]$ with success probability $\operatorname{Tr}[\mathcal{F}(\rho)]$. However, in the Heisenberg picture, the transformed measurement operators after conditioning on success are in general not valid POVMs (see recent discussions in Refs.~\cite{Arai2024arXiv,Ku2026npj}). For this reason, the robustness defined in Eq.~\eqref{eq:robustness JM} with an additional filtering operation acting on the measurement assemblage cannot be used directly.
    
    To overcome this difficulty, we adopt an operational definition of measurement incompatibility via quantum steerability \cite{WisemanPRL2007}. Quantum steering considers a scenario in which Alice and Bob share a bipartite state $\rho^{AB}$, and Alice performs a measurement from a measurement assemblage. Eventurally, this induces on a collection of Bob's conditional states $\sigma_{a|x}  = \operatorname{Tr_{A}} \left[ \rho^{AB} \left( A_{a|x} \otimes \mathbb{I} \right) \right]$, called a state assemblage $\bm{\sigma}=\{\sigma_{a|x}\}_{a,x}$.
    
    A state assemblage is steerable if it does not admit a local-hidden-state (LHS) model, i.e., if $\sigma_{a|x} \neq \sum_{\lambda} p(\lambda)p(a|x, \lambda)\rho_{\lambda}$. In this case, Bob's state is manipulated by Alice's measurement outcome, rather than being determined by a hidden variable $\lambda$. The set of all assemblages admitting an LHS model is denoted by $\mathbb{LHS}$.

    There exists a one-to-one correspondence between steering and measurement incompatibility~\cite{Quantino2014PRL, Uola2014PRL, Uola2015PRL}, captured by steering-equivalence-observable (SEO) measurement assemblage. For a state assemblage generated from a pure entangled state, the SEO is defined as $A_{a|x} \coloneqq \rho_{B}^{-\frac{1}{2}} \sigma_{a|x} \rho_{B}^{-\frac{1}{2}}$, where $\rho_{B} = \sum_{a}\sigma_{a|x}$ is Bob's reduced state. A state assemblage is steerable if and only if its SEO is incompatible. 
    The SEO mapping also allows us to transfer the action of a filter on measurement assemblages (in the Heisenberg picture) into its action on state assemblages (in the Schrodinger picture) via 
    $$\mathcal{F}(\bm{\sigma}):=\{\sigma'_{a|x}\}_{a,x}$$ with $\sigma'_{a|x}=\frac{\mathcal{F}(\sigma_{a|x})}{\operatorname{Tr}[\sum_{a}\mathcal{F}(\sigma_{a|x})]}~~\forall~~a,x.$
    On the other hand, the relationship between the steering and measurement incompatibility can be extended to their channel variants. Specifically, if a channel breaks quantum steerability for the maximally entangled state, its dual channel in the Heisenberg picture must break measurement incomparability for any measurements.
    
    In other words, if a channel is non-steerability breaking, it must preserve measurement incompatibility. Those facts make steering a suitable operational framework for analyzing the activation of the IA channel under quantum filters~\cite{Ku2026npj}.

    Inspired by the steering scenario, we can define another measurement incompatibility preservability robustness in the steering scenario. First, consider the maximally entangled state $\ket{\Phi^+}=\sum_i^d\frac{1}{\sqrt{d}}\ket{ii}$ as the initial state and let $\mathbf{M}$ be a measurement assemblage performed on Alice's subsystem. For a given channel $\mathcal{N}$, the corresponding state assemblage is defined as
    \begin{equation}
        \begin{aligned}
            \sigma_{\mathcal{N}, a|x} = \operatorname{Tr}_A \left[ (M_{a|x} \otimes \openone) (\mathcal{N} \otimes \mathcal{I} ) \ket{\Phi^+}\bra{\Phi^+}  \right], \quad \forall a, x.
        \end{aligned}
    \end{equation}
    Here, $\mathcal{I}$ denotes the identity channel on Bob's subsystem.
    Accordingly, for a arbitrary noise channel $\mathcal{W}$ in Eq.~\eqref{eq:optimization JM}, we define
    \begin{equation}
        \begin{aligned}
            \pi_{\mathcal{W}, a|x} = \operatorname{Tr}_A \left[ (M_{a|x} \otimes \openone) (\mathcal{W} \otimes \mathcal{I} ) \ket{\Phi^+}\bra{\Phi^+} \right], \quad \forall a, x
        \end{aligned}
    \end{equation}
    as the corresponding noise assemblage. Then, for a fixed measurement assemblage $\mathbf{M}$, the new measurement incompatibility preservability robustness can be defined by the following optimization problems as
    \begin{equation}
        \begin{aligned}
            &\mathcal{R}_{\text{IP}}^{\text{LHS}}(\mathcal{N}; \mathbf{M}) \coloneqq \\ &\min_{t, \mathcal{W} } \left\{t>=0 \left| \frac{\sigma_{\mathcal{N}, a|x} + t\pi_{\mathcal{W}, a|x}}{1+t}=\sigma^{\text{LHS}}_{a|x}, \quad \forall a, x \right\} \right..
            \label{eq:optimization LHS}
        \end{aligned}
    \end{equation}
    where \(\left\{ \sigma^{\text{LHS}}_{a|x} \right\}_{a, x} \in \mathbb{LHS}\). Similarly, we further optimize the measurement assemblage and define
    \begin{equation}
        \begin{aligned}
            \mathcal{R}^{\text{LHS}}_{\text{IP}}(\mathcal{N}) \coloneqq \max_{\mathbf{M}} \mathcal{R}^{\text{LHS}}_{\text{IP}}(\mathcal{N}; \mathbf{M}).
            \label{eq:robustness LHS}
        \end{aligned}
    \end{equation}
    We now have our second main result:
    
    \begin{result}
        For any set of measurements \(\mathbf{M}:=\{M_{a|x}\}_{a,x}\) and any quantum channel \(\mathcal{N}\), we have 
        \(
        \mathcal{R}^{\text{JM}}_{\text{IP}}(\mathcal{N};\mathbf{M})
        =
        \mathcal{R}^{\text{LHS}}_{\text{IP}}(\mathcal{N};\mathbf{M}).
        \)
        Consequently,
        \(
        \mathcal{R}^{\text{JM}}_{\text{IP}}(\mathcal{N})
        =
        \mathcal{R}^{\text{LHS}}_{\text{IP}}(\mathcal{N}).
        \)
        \label{result:equality}
    \end{result}
    \begin{proof}
        We prove the result by showing both the two inequalities
        \(
            \mathcal{R}^{\text{JM}}_{\text{IP}}(\mathcal{N};\mathbf{M}) \geq \mathcal{R}^{\text{LHS}}_{\text{IP}}(\mathcal{N};\mathbf{M})
        \)
        and
        \(
            \mathcal{R}^{\text{LHS}}_{\text{IP}}(\mathcal{N};\mathbf{M})
            \geq
            \mathcal{R}^{\text{JM}}_{\text{IP}}(\mathcal{N};\mathbf{M})
        \)
        holds.
        %For simplicity, we assume that the optima are attained.

        We start with recalling the following useful facts: for any operator \(X\) and the maximally entangled state $\ket{\Phi^+}$ we have,
        \begin{equation}
            \begin{aligned}
                \operatorname{Tr}_A
                \left[
                (X\otimes \openone)
                \ket{\Phi^+}\bra{\Phi^+}
                \right]
                =
                \frac{X^T}{d},
                \label{eq:MES identity-1}
            \end{aligned}
        \end{equation}
        where the transpose is taken with respect to the basis defining \(\ket{\Phi^+}\). Therefore,
        \begin{equation}
            \begin{aligned}
                \sigma_{\mathcal{N},a|x}
                =
                \frac{[\mathcal{N}^{\dagger}(M_{a|x})]^T}{d},
                \qquad
                \pi_{\mathcal{W},a|x}
                =
                \frac{[\mathcal{W}^{\dagger}(M_{a|x})]^T}{d}.
                \label{eq:MES identity-2}
            \end{aligned}
        \end{equation}
        Moreover, since \(\mathcal{N}^{\dagger}\) and \(\mathcal{W}^{\dagger}\) are unital and
        \(\sum_a M_{a|x}=\openone\), we have
        \begin{equation}
            \begin{aligned}
                \sum_a\sigma_{\mathcal{N},a|x}
                =
                \sum_a\pi_{\mathcal{W},a|x}
                =
                \frac{\openone}{d}.
                \label{eq:MES identity-3}
            \end{aligned}
        \end{equation}
        
        Now, we start with proving
        \[
            \mathcal{R}^{\text{JM}}_{\text{IP}}(\mathcal{N};\mathbf{M})
            \geq
            \mathcal{R}^{\text{LHS}}_{\text{IP}}(\mathcal{N};\mathbf{M}).
        \]
        Let \(t^*\) and \(\mathcal{W}^*\) be an optimal solution of
        \(\mathcal{R}^{\text{JM}}_{\text{IP}}(\mathcal{N};\mathbf{M})\). By definition in Eq.~\eqref{eq:optimization JM},
        \[
            \widetilde{M}_{a|x}
            \coloneqq
            \frac{
            \mathcal{N}^{\dagger}
            +
            t^*\mathcal{W}^{*\dagger}
            }{1+t^*}
            (M_{a|x})
        \]
        forms a jointly measurable measurement assemblage. %That is,
        %\[
        %    \{\widetilde{M}_{a|x}\}_{a,x}\in\mathbb{JM}.
        %\]
       Consider the assemblage obtained by measuring one subsystem of the maximally entangled state with \(\{\widetilde{M}_{a|x}\}_{a,x}\):
        \[
            \widetilde{\sigma}_{a|x}
            \coloneqq
            \operatorname{Tr}_A
            \left[
            (\widetilde{M}_{a|x}\otimes \openone)
            \ket{\Phi^+}\bra{\Phi^+}
            \right].
        \]
        Since \(\{\widetilde{M}_{a|x}\}_{a,x}\) is jointly measurable, the assemblage
        \(\{\widetilde{\sigma}_{a|x}\}_{a,x}\) is unsteerable.%, i.e.,
        %\[
        %    \{\widetilde{\sigma}_{a|x}\}_{a,x}\in\mathbb{LHS}.
        %\]
        Using the Eq.~\eqref{eq:MES identity-1}, ~\eqref{eq:MES identity-2} and the linearity of the transpose, we obtain
        \[
            \begin{aligned}
                \widetilde{\sigma}_{a|x}
                &=
                \frac{
                \left[
                \frac{
                \mathcal{N}^{\dagger}
                +
                t^*\mathcal{W}^{*\dagger}
                }{1+t^*}
                (M_{a|x})
                \right]^T
                }{d} \\
                &=
                \frac{
                [\mathcal{N}^{\dagger}(M_{a|x})]^T
                +
                t^*[\mathcal{W}^{*\dagger}(M_{a|x})]^T
                }{d(1+t^*)} \\
                &=
                \frac{
                \sigma_{\mathcal{N},a|x}
                +
                t^*\pi_{\mathcal{W}^*,a|x}
                }{1+t^*}.
            \end{aligned}
        \]
        Therefore, the same \(t^*\) and \(\mathcal{W}^*\) are feasible for
        \(\mathcal{R}^{\text{LHS}}_{\text{IP}}(\mathcal{N};\mathbf{M})\). Hence,
        \[
            \mathcal{R}^{\text{JM}}_{\text{IP}}(\mathcal{N};\mathbf{M})
            = t^* \geq
            \mathcal{R}^{\text{LHS}}_{\text{IP}}(\mathcal{N};\mathbf{M}).
        \]
        Next, we prove the opposite direction
        \[
            \mathcal{R}^{\text{LHS}}_{\text{IP}}(\mathcal{N};\mathbf{M})
            \geq
            \mathcal{R}^{\text{JM}}_{\text{IP}}(\mathcal{N};\mathbf{M}).
        \]
        Let \(t^*\) and \(\mathcal{W}^*\) be an optimal solution of
        \(\mathcal{R}^{\text{LHS}}_{\text{IP}}(\mathcal{N};\mathbf{M})\), and defining
        \[
            \widetilde{\sigma}_{a|x}
            \coloneqq
            \frac{
            \sigma_{\mathcal{N},a|x}
            +
            t^*\pi_{\mathcal{W}^*,a|x}
            }{1+t^*}
        \]
        is an LHS assemblage. %That is,
        %\[
        %    \{\widetilde{\sigma}_{a|x}\}_{a,x}\in\mathbb{LHS}.
        %\]
        With the Eq.~\eqref{eq:MES identity-3}
        \[
            \sum_a\widetilde{\sigma}_{a|x} = \frac{\openone}{d} = \rho_B
        \]
        By the definition of SEO mentioned before, the SEO of \(\{\widetilde{\sigma}_{a|x}\}_{a,x}\) are
        \[
            \widetilde{A}_{a|x}
            =
            \left(\frac{\openone}{d}\right)^{-\frac{1}{2}}
            \widetilde{\sigma}_{a|x}
            \left(\frac{\openone}{d}\right)^{-\frac{1}{2}}
            =
            d\widetilde{\sigma}_{a|x},
        \]
        which is jointly measurable. Using the Eq.~\eqref{eq:MES identity-2} to rewrite \(\widetilde{\sigma}_{a|x}\), we obtain
        \[
            \begin{aligned}
                &\widetilde{\sigma}_{a|x}
                =
                \frac{
                [\mathcal{N}^{\dagger}(M_{a|x})]^T
                +
                t^*[\mathcal{W}^{*\dagger}(M_{a|x})]^T
                }{d(1+t^*)} \\
                &\implies
                d\widetilde{\sigma}_{a|x}
                =
                \frac{
                [\mathcal{N}^{\dagger}(M_{a|x})]^T
                +
                t^*[\mathcal{W}^{*\dagger}(M_{a|x})]^T
                }{(1+t^*)} = \widetilde{A}_{a|x} \\
                & \implies
                \frac{
                \mathcal{N}^{\dagger}(M_{a|x})
                +
                t^*\mathcal{W}^{*\dagger}(M_{a|x})
                }{(1+t^*)}
                = \widetilde{A}_{a|x}^T
            \end{aligned}
        \]
        Since transposition preserves joint measurability, \(t^*\) and \(\mathcal{W}^*\) must be feasible for
        \(\mathcal{R}^{\text{JM}}_{\text{IP}}(\mathcal{N};\mathbf{M})\). Hence,
        \[
            \mathcal{R}^{\text{LHS}}_{\text{IP}}(\mathcal{N};\mathbf{M})
            = t^* \geq
            \mathcal{R}^{\text{JM}}_{\text{IP}}(\mathcal{N};\mathbf{M}).
        \]
        
        Combining the two inequalities, we conclude that
        \[
            \mathcal{R}^{\text{JM}}_{\text{IP}}(\mathcal{N};\mathbf{M})
            =
            \mathcal{R}^{\text{LHS}}_{\text{IP}}(\mathcal{N};\mathbf{M}).
        \]
        Since this equality holds for every measurement set \(\mathbf{M}\), maximizing both sides over \(\mathbf{M}\) gives
        \[
            \mathcal{R}^{\text{JM}}_{\text{IP}}(\mathcal{N})
            =
            \mathcal{R}^{\text{LHS}}_{\text{IP}}(\mathcal{N}),
        \] which conclude the proof.
    \end{proof}
    
    %Since \(\mathcal{R}^{\text{JM}}_{\text{IP}}(\mathcal{N})=\mathcal{R}^{\text{LHS}}_{\text{IP}}(\mathcal{N})\), we can naturally found that \(\mathcal{R}^{\text{LHS}}_{\text{IP}}(\mathcal{N})\) is also a valid resource monotone that satisfies the properties in \eqref{eq:resource monotone}.

    The direct consequence of the Result~\ref{result:equality} is that \(\mathcal{R}^{\text{LHS}}_{\text{IP}}(\mathcal{N})\) is also a valid incompatiblity preservability monotone that satisfies the properties in Eq.~\eqref{eq:resource monotone}.
    
    \section{No-go of activation of IA channel via quantum pre-processing}

    In Result~\ref{result:equality}, we have shown how to use the SEO to convert the two measurement incompatibility preservability robustness defined through measurement incomparability and steering.
    We now present our third main result, which takes the form of a no-go theorem:
    \begin{result}
    An IA channel cannot be operationally activated by pre-filtering. Specifically, no filtering operation applied before an IA channel can enable the generation of quantum steerability.
    \label{result:no-go}
    \end{result}

    \begin{proof}
    Within the resource theory of measurement incompatibility preservability~\cite{Hsieh2025PRA}, the free operation is characterized through $\mathcal{N} \mapsto \sum_{k} \mathcal{D}_k  \circ \mathcal{N} \circ \mathcal{F}_k$, where $D_k$ and $F_k$ are CP trace preserving and CP-TNI with classical communication $k$.
    By definition of the free operation, the post quantum channel $D_k$ is impossible to activate in an IA channel. Thus, we consider the pre-filtering operation $\mathcal{F}$ acting on the given IA channel. Due to the equivalent relation between incompatibility annihilating and steerability breaking channel~\cite{KuPRXQ2022,Hsieh2025PRA}, we can reformulate the activation problem as finding the optimal filtering operation followed by an IA channel with the maximally entangled state $\ket{\Phi^+}=\sum_i^d\frac{1}{\sqrt{d}}\ket{ii}$ as the input such that the output state is steerable.
    
    Mathematically, the output state can be expressed as 
    \[
        \rho=\frac{ \mathcal{N}\circ \mathcal{F}\otimes \openone \ket{\Phi^+}\bra{\Phi^+}}{\text{Tr}[\mathcal{N}\circ \mathcal{F}\otimes \openone \ket{\Phi^+}\bra{\Phi^+}]}.
    \]
    We now apply Eq.~\eqref{eq:MES identity-1}, and, thus, the output state can be reformulated as
    \[
        \rho=\frac{ \mathcal{N}\otimes \mathcal{F}' \ket{\Phi^+}\bra{\Phi^+}}{\text{Tr}[\mathcal{N}\circ \mathcal{F}\otimes \openone \ket{\Phi^+}\bra{\Phi^+}]}
    \]
    where $\mathcal{F}'$ is the effective filtering operation acting on Bob's side. This can be seen as applying a local filtering operation on the steering scenario. Thus, we adopt the Theorem 1 in Refs.~\cite{Ku2022NC,Wang2024} which explicit states that the filtering operation acting on Bob's side cannot activate quantum steerability, which completes the proof.
    \end{proof}

    \section{Activation of IA channel}    
    \begin{figure*}[t]
        \centering
    
        \newlength{\panelw}
        \setlength{\panelw}{0.292\textwidth}
    
        \newlength{\cbarw}
        \setlength{\cbarw}{0.055\textwidth}
    
        \setlength{\tabcolsep}{0pt}
        \renewcommand{\arraystretch}{0}
    
        \begin{tabular}{@{}p{\panelw}@{\hspace{0.012\textwidth}}
                        p{\panelw}@{\hspace{0.012\textwidth}}
                        p{\panelw}@{\hspace{0.006\textwidth}}
                        p{\cbarw}@{}}
    
            % ---------- first row: (a)--(c) ----------
            \subfloat[]{%
                \includegraphics[width=\linewidth,page=1]{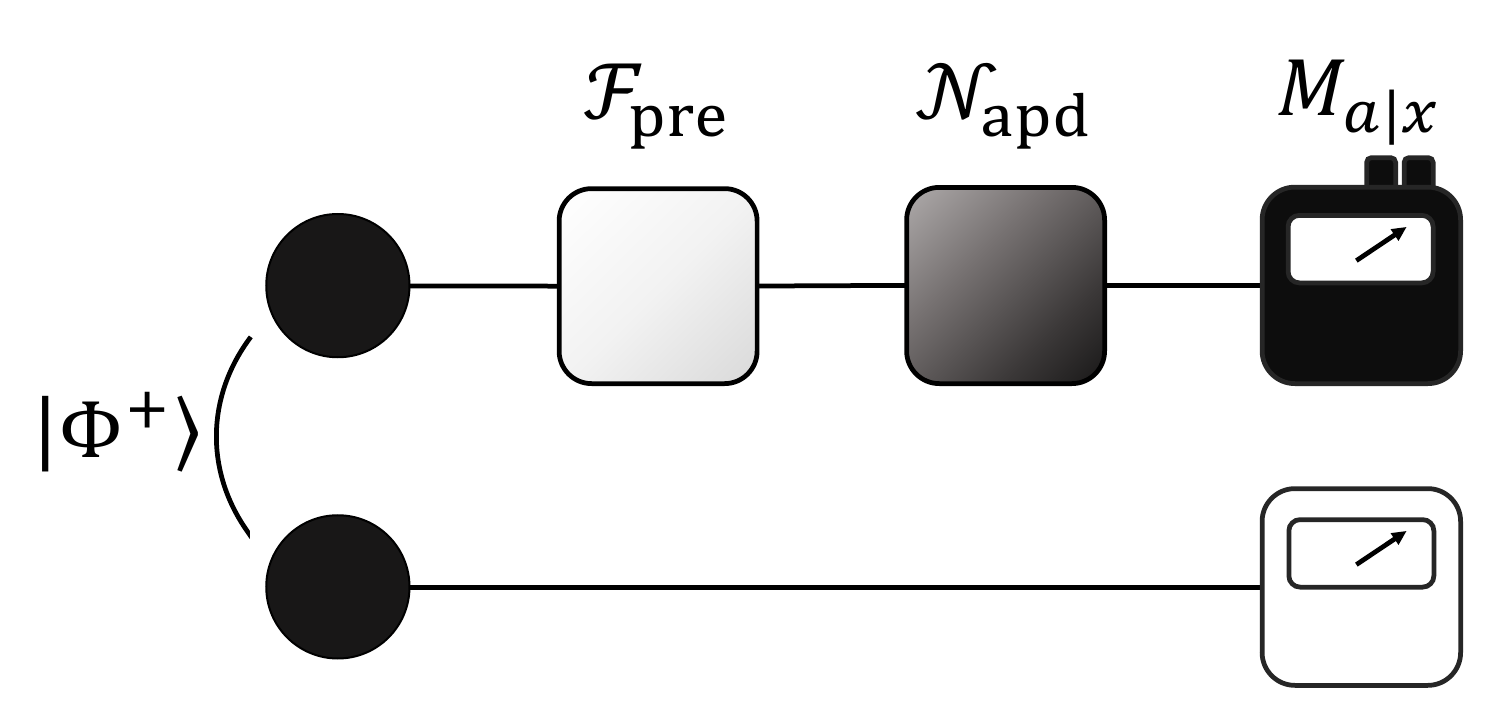}
                \label{fig:pre_filter_setup}
            }
            &
            \subfloat[]{%
                \includegraphics[width=\linewidth,page=2]{figures/setup_figures.pdf}
                \label{fig:post_filter_setup}
            }
            &
            \subfloat[]{%
                \includegraphics[width=\linewidth,page=3]{figures/setup_figures.pdf}
                \label{fig:prepost_filter_setup}
            }
            &
            \\[0.8em]
    
            % ---------- second row: (d)--(f) + colorbar ----------
            \subfloat[]{%
                \includegraphics[width=\linewidth]{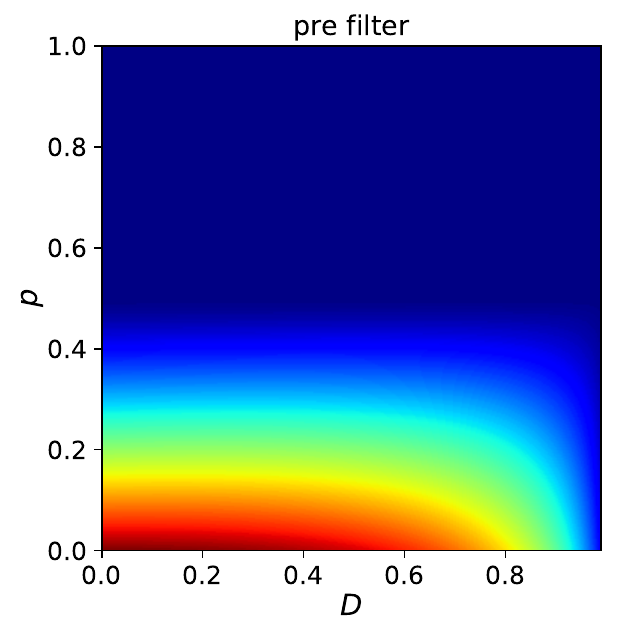}
                \label{fig:pre only result}
            }
            &
            \subfloat[]{%
                \includegraphics[width=\linewidth]{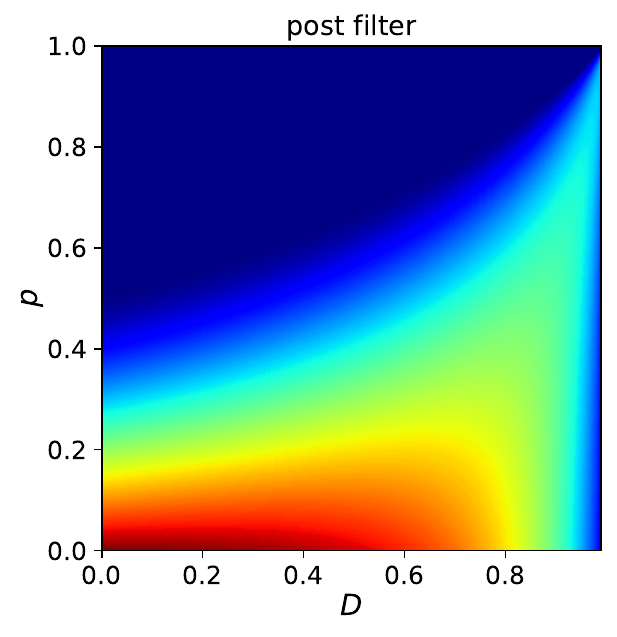}
                \label{fig:post only result}
            }
            &
            \subfloat[]{%
                \includegraphics[width=\linewidth]{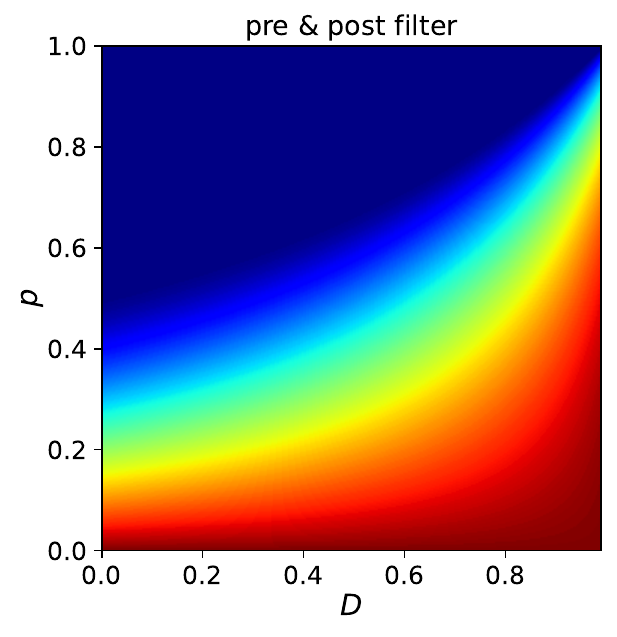}
                \label{fig:prepost result}
            }
            &
            \raisebox{-1.125\height}{%
                \includegraphics[height=1.0\panelw]{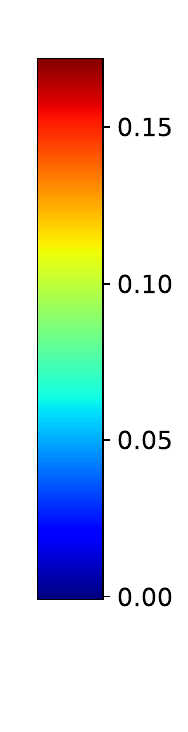}
            }
    
        \end{tabular}

        \caption{
        Schematic diagrams and numerical results for different filtering scenarios.
        (a)--(c) show the setups for the pre-filter only, post-filter only, and pre-post-filter cases, respectively.
        And (d)--(f) show the corresponding numerical results.
        }
        \label{fig:big figure}
    \end{figure*}
 
    With the robustness quantification and no-go theorem, we can now demonstrate the activation of IA channels via a filtering operation. As a concrete demonstration, we consider the activation of the IA amplitude-damping channels. The rule of the amplitude-damping channel is analogous to the spontaneous emission process in the sense that it transmits the excited state $\ket{1}$ to ground state $\ket{0}$ with probability $p$, while preserving $\ket{0}$ with certainty, namely
    \begin{equation*}
        \begin{aligned}
        &\operatorname{\mathcal{N}_{\text{adc}}} \left( \ket{0} \right) = \ket{0}, \\ &\operatorname{\mathcal{N}_{\text{adc}}} \left( \ket{1} \right) = p\ket{0} + (1-p)\ket{1}.
        \end{aligned}
    \end{equation*}
    Consequently, the output state under the amplitude-damping channel is expressed as
    $\mathcal{N}_{\text{adc}}(\rho) = (1 - p)\rho + p \ket{0}\bra{0}$.
    The amplitude-damping channel preserves measurement incompatibility with the two-measurement setting if $p\leq 0.5$~\cite{KuPRA2023,ChangPRR2024}. Thus, the activation of IA channels can be sufficiently demonstrated by showing that the robustness becomes nonzero after inserting the post-filtering operation for $p\leq 0.5$ (see Fig.~\ref{fig:big figure} (b)). Indeed, once we apply the post-filtering operation $\mathcal{F}_{post}(\rho)=K\rho K^{\dagger}$ with Kraus operator 
    \begin{eqnarray}
        \begin{aligned}
            K =
            \begin{bmatrix}
                \sqrt{1-D} &0 \\
                0 &1
            \end{bmatrix},
        \end{aligned}
        \label{eq:post filter}
    \end{eqnarray}
    after the amplitude-damping channel when $p\leq 0.5$, we can observe the activation of the IA channel. We numerically present the \(\mathcal{R}^{\text{LHS}}_{\text{IP}}(\mathcal{N})\) over $p$ and $D$ in Fig.~\ref{fig:big figure} (d), where $D$ is parameter of filters. Note that the measurement assemblage $\{M_{a|x}\}_{a,x}$ is obtained via the see-saw algorithm.
    
    In Result~\ref{result:no-go}, it has been shown that any pre-filtering operation cannot be used to activate an IA channel (see also an example in Figs.~\ref{fig:big figure} (a) and (d) with pre-filtering operation introduced later). Nevertheless, one can combine both pre- and post-filtering operations to not only activate the IA channel but also distill it. %In Fig.~\ref{fig:big figure}, we present an explicit example. 
    Specifically, we first apply the same post-filtering operation in Eq.~\eqref{eq:post filter} on the amplitude-damping channel to activate its incompatibility annihilating properties when $p\leq 0.5$. On the other hand, the pre-filtering  operation $\mathcal{F}_{pre}(\rho)$ with Kraus operator
    \begin{equation}
        \begin{aligned}
            K=
            \begin{bmatrix}
                1 &0 \\
                0 &\sqrt{1-D}
            \end{bmatrix}
            \label{eq:pre filter}
        \end{aligned}
    \end{equation}
    is used for distilling robustness of steering-based measurement incompatibility preservability. In Fig.\ref{fig:big figure} (c) and (d), one indeed observes that the amplitude-damping IA channel can be activated and distilled via post- and pre-filtering operations, respectively.

\section{Conclusions}
\label{sec:conclusions}
    To study how to activate an IA channel, we introduce a computable measurement incompatibility preservability monotone \(\mathcal{R}^{\text{JM}}_{\text{IP}}\) under the resource theory of measurement incompatibility preservability. This new monotone is used as our figure-of-merit since the original definition of incompatibility-preservability robustness in Eq.~\eqref{eq:robustness IA} cannot be computed.

    We further established an equivalent steering-based monotone \(\mathcal{R}^{\text{LHS}}_{\text{IP}}\) in the Schr\"odinger picture, which overcomes the impossibility of a complete POVM after applying filtering operations in the Heisenberg picture. Based on this framework, we proved a strong no-go theorem showing that IA channels cannot be operationally activated by pre-filtering alone. In contrast, our numerical results demonstrate that IA activation is possible through post-filtering operations. Moreover, by combining pre- and post-filtering operations, both activation and distillation of IA channels can be realized in the same experiment.

    We conclude with a few open questions. In Refs.~\cite{Ku2022NC,Ku2026npj}, the optimal local filtering operations for a series of steering robustness are characterized to achieve the equivalent relation with the corresponding measurement incompatibility robustness. Here, we adopt the pre- and post-filtering operation in Eqs.~\eqref{eq:pre filter} and~\eqref{eq:post filter} for demonstrating IA activation and distillation. How to obtain the optimal filtering operations in the dynamical frameworks of quantum steering and measurement incompatibility?
    
    In addition, we apply a see-saw algorithm to potentially find out the optimal measurement assemblage in Eq.~\eqref{eq:robustness JM}. The same approach is also applied to finding the most incompatible measurements for achieving the optimal steerability in the level of quantum state~\cite{JessicaPRA2017}. A similar problem, but in a different scheme, has been solved. In quantifying a non-entanglement-breaking channel, the optimal input state is always the maximally entangled state. How can we bypass the optimization of measurement assemblage in Eq.~\eqref{eq:robustness JM}?

    %\begin{equation}
    %\frac{p^N_f}{N+1}\Lambda(\rho)+\frac{Np^{N-1}_f}{N+1}\sum_{ij}\bra{\phi_a}K_i\ket{\xi_f}\bra{\xi_f}K^\dagger_j\ket{\phi_i}K_i
    %\rho K^\dagger_j,
%\end{equation}
%where $p_f=|\sum_i\bra{\phi_a}K_i\ket{\xi_f}|^2$ is the probability of projecting the ancillary state after noisy channel $\sum_i K_i\ket{\phi_a}$ on $\ket{\xi_f}$. 

%\begin{result}
%Given a Kraus representation $\{K_i\}_i$ of the channel and a coefficient $\vec{\alpha}$ that defines the effective operator $T = \sum_i \alpha_i^ K_i$, then for any equivalent Kraus representation $\{K'_i\}_i$, there exists a $\vec{\alpha}'$ that generates the same operator $T$, i.e., $\sum_i \alpha_i^ K_i = \sum_i {\alpha'_i}^ K'_i$.
%\end{result}

%\begin{widetext}
%\begin{equation}
%= \left( \frac{\Lambda(\rho) + (N-1) F \rho F^\dagger}{N}\otimes \ket{+}\bra{+}\right)  + \left( \frac{\Lambda(\rho) - F\rho F^\dagger}{N}\otimes (\openone-\ket{+}\bra{+})\right).
%\end{equation}
%\end{widetext}

\section*{Acknowledgements} \label{sec:acknowledgements}
    The authors acknowledge fruitful discussions with Chung-Yun Hsieh, and Benjamin Stratton.
    This project is supported by National Science and Technology Council (NSTC) with Grant No. NSTC 115-2112-M-003 -010. F.N. is supported in part by: the Japan Science and Technology Agency (JST) [via the CREST Quantum Frontiers program Grant No. JPMJCR24I2, the Quantum Leap Flagship Program (Q-LEAP), and the Moonshot R\&D Grant Number JPMJMS2061].
    All results are figured out, obtained, and verified by the authors. We used an LLM (and Grammarly) only to improve English writing, to check grammar, and to fix typos.

%\bibliography{ref.bib}

\begin{thebibliography}{40}%
\makeatletter
\providecommand \@ifxundefined [1]{%
 \@ifx{#1\undefined}
}%
\providecommand \@ifnum [1]{%
 \ifnum #1\expandafter \@firstoftwo
 \else \expandafter \@secondoftwo
 \fi
}%
\providecommand \@ifx [1]{%
 \ifx #1\expandafter \@firstoftwo
 \else \expandafter \@secondoftwo
 \fi
}%
\providecommand \natexlab [1]{#1}%
\providecommand \enquote  [1]{``#1''}%
\providecommand \bibnamefont  [1]{#1}%
\providecommand \bibfnamefont [1]{#1}%
\providecommand \citenamefont [1]{#1}%
\providecommand \href@noop [0]{\@secondoftwo}%
\providecommand \href [0]{\begingroup \@sanitize@url \@href}%
\providecommand \@href[1]{\@@startlink{#1}\@@href}%
\providecommand \@@href[1]{\endgroup#1\@@endlink}%
\providecommand \@sanitize@url [0]{\catcode `\\12\catcode `\$12\catcode `\&12\catcode `\#12\catcode `\^12\catcode `\_12\catcode `\%12\relax}%
\providecommand \@@startlink[1]{}%
\providecommand \@@endlink[0]{}%
\providecommand \url  [0]{\begingroup\@sanitize@url \@url }%
\providecommand \@url [1]{\endgroup\@href {#1}{\urlprefix }}%
\providecommand \urlprefix  [0]{URL }%
\providecommand \Eprint [0]{\href }%
\providecommand \doibase [0]{https://doi.org/}%
\providecommand \selectlanguage [0]{\@gobble}%
\providecommand \bibinfo  [0]{\@secondoftwo}%
\providecommand \bibfield  [0]{\@secondoftwo}%
\providecommand \translation [1]{[#1]}%
\providecommand \BibitemOpen [0]{}%
\providecommand \bibitemStop [0]{}%
\providecommand \bibitemNoStop [0]{.\EOS\space}%
\providecommand \EOS [0]{\spacefactor3000\relax}%
\providecommand \BibitemShut  [1]{\csname bibitem#1\endcsname}%
\let\auto@bib@innerbib\@empty
%</preamble>
\bibitem [{\citenamefont {Busch}\ \emph {et~al.}(2014)\citenamefont {Busch}, \citenamefont {Lahti},\ and\ \citenamefont {Werner}}]{BuschRMP}%
  \BibitemOpen
  \bibfield  {author} {\bibinfo {author} {\bibfnamefont {P.}~\bibnamefont {Busch}}, \bibinfo {author} {\bibfnamefont {P.}~\bibnamefont {Lahti}},\ and\ \bibinfo {author} {\bibfnamefont {R.~F.}\ \bibnamefont {Werner}},\ }\bibfield  {title} {\bibinfo {title} {Colloquium: Quantum root-mean-square error and measurement uncertainty relations},\ }\href {https://doi.org/10.1103/RevModPhys.86.1261} {\bibfield  {journal} {\bibinfo  {journal} {Rev. Mod. Phys.}\ }\textbf {\bibinfo {volume} {86}},\ \bibinfo {pages} {1261} (\bibinfo {year} {2014})}\BibitemShut {NoStop}%
\bibitem [{\citenamefont {Heinosaari}\ \emph {et~al.}(2016)\citenamefont {Heinosaari}, \citenamefont {Miyadera},\ and\ \citenamefont {Ziman}}]{Heinosaari_2016}%
  \BibitemOpen
  \bibfield  {author} {\bibinfo {author} {\bibfnamefont {T.}~\bibnamefont {Heinosaari}}, \bibinfo {author} {\bibfnamefont {T.}~\bibnamefont {Miyadera}},\ and\ \bibinfo {author} {\bibfnamefont {M.}~\bibnamefont {Ziman}},\ }\bibfield  {title} {\bibinfo {title} {An invitation to quantum incompatibility},\ }\href {https://doi.org/10.1088/1751-8113/49/12/123001} {\bibfield  {journal} {\bibinfo  {journal} {Journal of Physics A: Mathematical and Theoretical}\ }\textbf {\bibinfo {volume} {49}},\ \bibinfo {pages} {123001} (\bibinfo {year} {2016})}\BibitemShut {NoStop}%
\bibitem [{\citenamefont {Guhne}\ \emph {et~al.}(2023)\citenamefont {Guhne}, \citenamefont {Haapasalo}, \citenamefont {Kraft}, \citenamefont {Pellonp\"a\"a},\ and\ \citenamefont {Uola}}]{GuhneRMP2023}%
  \BibitemOpen
  \bibfield  {author} {\bibinfo {author} {\bibfnamefont {O.}~\bibnamefont {Guhne}}, \bibinfo {author} {\bibfnamefont {E.}~\bibnamefont {Haapasalo}}, \bibinfo {author} {\bibfnamefont {T.}~\bibnamefont {Kraft}}, \bibinfo {author} {\bibfnamefont {J.-P.}\ \bibnamefont {Pellonp\"a\"a}},\ and\ \bibinfo {author} {\bibfnamefont {R.}~\bibnamefont {Uola}},\ }\bibfield  {title} {\bibinfo {title} {Colloquium: Incompatible measurements in quantum information science},\ }\href {https://doi.org/10.1103/RevModPhys.95.011003} {\bibfield  {journal} {\bibinfo  {journal} {Rev. Mod. Phys.}\ }\textbf {\bibinfo {volume} {95}},\ \bibinfo {pages} {011003} (\bibinfo {year} {2023})}\BibitemShut {NoStop}%
\bibitem [{\citenamefont {Brunner}\ \emph {et~al.}(2014)\citenamefont {Brunner}, \citenamefont {Cavalcanti}, \citenamefont {Pironio}, \citenamefont {Scarani},\ and\ \citenamefont {Wehner}}]{BrunnerRMP}%
  \BibitemOpen
  \bibfield  {author} {\bibinfo {author} {\bibfnamefont {N.}~\bibnamefont {Brunner}}, \bibinfo {author} {\bibfnamefont {D.}~\bibnamefont {Cavalcanti}}, \bibinfo {author} {\bibfnamefont {S.}~\bibnamefont {Pironio}}, \bibinfo {author} {\bibfnamefont {V.}~\bibnamefont {Scarani}},\ and\ \bibinfo {author} {\bibfnamefont {S.}~\bibnamefont {Wehner}},\ }\bibfield  {title} {\bibinfo {title} {Bell nonlocality},\ }\href {https://doi.org/10.1103/RevModPhys.86.419} {\bibfield  {journal} {\bibinfo  {journal} {Rev. Mod. Phys.}\ }\textbf {\bibinfo {volume} {86}},\ \bibinfo {pages} {419} (\bibinfo {year} {2014})}\BibitemShut {NoStop}%
\bibitem [{\citenamefont {Quintino}\ \emph {et~al.}(2014{\natexlab{a}})\citenamefont {Quintino}, \citenamefont {V\'ertesi},\ and\ \citenamefont {Brunner}}]{QuintinoPRL2014}%
  \BibitemOpen
  \bibfield  {author} {\bibinfo {author} {\bibfnamefont {M.~T.}\ \bibnamefont {Quintino}}, \bibinfo {author} {\bibfnamefont {T.}~\bibnamefont {V\'ertesi}},\ and\ \bibinfo {author} {\bibfnamefont {N.}~\bibnamefont {Brunner}},\ }\bibfield  {title} {\bibinfo {title} {Joint measurability, einstein-podolsky-rosen steering, and bell nonlocality},\ }\href {https://doi.org/10.1103/PhysRevLett.113.160402} {\bibfield  {journal} {\bibinfo  {journal} {Phys. Rev. Lett.}\ }\textbf {\bibinfo {volume} {113}},\ \bibinfo {pages} {160402} (\bibinfo {year} {2014}{\natexlab{a}})}\BibitemShut {NoStop}%
\bibitem [{\citenamefont {Chen}\ \emph {et~al.}(2016)\citenamefont {Chen}, \citenamefont {Budroni}, \citenamefont {Liang},\ and\ \citenamefont {Chen}}]{ChenPRL2016}%
  \BibitemOpen
  \bibfield  {author} {\bibinfo {author} {\bibfnamefont {S.-L.}\ \bibnamefont {Chen}}, \bibinfo {author} {\bibfnamefont {C.}~\bibnamefont {Budroni}}, \bibinfo {author} {\bibfnamefont {Y.-C.}\ \bibnamefont {Liang}},\ and\ \bibinfo {author} {\bibfnamefont {Y.-N.}\ \bibnamefont {Chen}},\ }\bibfield  {title} {\bibinfo {title} {Natural framework for device-independent quantification of quantum steerability, measurement incompatibility, and self-testing},\ }\href {https://doi.org/10.1103/PhysRevLett.116.240401} {\bibfield  {journal} {\bibinfo  {journal} {Phys. Rev. Lett.}\ }\textbf {\bibinfo {volume} {116}},\ \bibinfo {pages} {240401} (\bibinfo {year} {2016})}\BibitemShut {NoStop}%
\bibitem [{\citenamefont {Cavalcanti}\ and\ \citenamefont {Skrzypczyk}(2016)}]{CavalcantiPRA2016}%
  \BibitemOpen
  \bibfield  {author} {\bibinfo {author} {\bibfnamefont {D.}~\bibnamefont {Cavalcanti}}\ and\ \bibinfo {author} {\bibfnamefont {P.}~\bibnamefont {Skrzypczyk}},\ }\bibfield  {title} {\bibinfo {title} {Quantitative relations between measurement incompatibility, quantum steering, and nonlocality},\ }\href {https://doi.org/10.1103/PhysRevA.93.052112} {\bibfield  {journal} {\bibinfo  {journal} {Phys. Rev. A}\ }\textbf {\bibinfo {volume} {93}},\ \bibinfo {pages} {052112} (\bibinfo {year} {2016})}\BibitemShut {NoStop}%
\bibitem [{\citenamefont {Ducuara}\ \emph {et~al.}(2026)\citenamefont {Ducuara}, \citenamefont {Lipka-Bartosik}, \citenamefont {Susa},\ and\ \citenamefont {Skrzypczyk}}]{Ducuara2026PRL}%
  \BibitemOpen
  \bibfield  {author} {\bibinfo {author} {\bibfnamefont {A.~F.}\ \bibnamefont {Ducuara}}, \bibinfo {author} {\bibfnamefont {P.}~\bibnamefont {Lipka-Bartosik}}, \bibinfo {author} {\bibfnamefont {C.~E.}\ \bibnamefont {Susa}},\ and\ \bibinfo {author} {\bibfnamefont {P.}~\bibnamefont {Skrzypczyk}},\ }\bibfield  {title} {\bibinfo {title} {All incompatible sets of measurements can generate nonlocality using quantum inputs},\ }\href {https://doi.org/10.1103/fgh1-gq8p} {\bibfield  {journal} {\bibinfo  {journal} {Phys. Rev. Lett.}\ }\textbf {\bibinfo {volume} {136}},\ \bibinfo {pages} {120205} (\bibinfo {year} {2026})}\BibitemShut {NoStop}%
\bibitem [{\citenamefont {Hsieh}\ \emph {et~al.}(2016)\citenamefont {Hsieh}, \citenamefont {Liang},\ and\ \citenamefont {Lee}}]{Hsieh2016PRA}%
  \BibitemOpen
  \bibfield  {author} {\bibinfo {author} {\bibfnamefont {C.-Y.}\ \bibnamefont {Hsieh}}, \bibinfo {author} {\bibfnamefont {Y.-C.}\ \bibnamefont {Liang}},\ and\ \bibinfo {author} {\bibfnamefont {R.-K.}\ \bibnamefont {Lee}},\ }\bibfield  {title} {\bibinfo {title} {Quantum steerability: Characterization, quantification, superactivation, and unbounded amplification},\ }\href {https://doi.org/10.1103/PhysRevA.94.062120} {\bibfield  {journal} {\bibinfo  {journal} {Phys. Rev. A}\ }\textbf {\bibinfo {volume} {94}},\ \bibinfo {pages} {062120} (\bibinfo {year} {2016})}\BibitemShut {NoStop}%
\bibitem [{\citenamefont {Uola}\ \emph {et~al.}(2020)\citenamefont {Uola}, \citenamefont {Costa}, \citenamefont {Nguyen},\ and\ \citenamefont {G\"uhne}}]{Uola2020RMP}%
  \BibitemOpen
  \bibfield  {author} {\bibinfo {author} {\bibfnamefont {R.}~\bibnamefont {Uola}}, \bibinfo {author} {\bibfnamefont {A.~C.~S.}\ \bibnamefont {Costa}}, \bibinfo {author} {\bibfnamefont {H.~C.}\ \bibnamefont {Nguyen}},\ and\ \bibinfo {author} {\bibfnamefont {O.}~\bibnamefont {G\"uhne}},\ }\bibfield  {title} {\bibinfo {title} {Quantum steering},\ }\href {https://doi.org/10.1103/RevModPhys.92.015001} {\bibfield  {journal} {\bibinfo  {journal} {Rev. Mod. Phys.}\ }\textbf {\bibinfo {volume} {92}},\ \bibinfo {pages} {015001} (\bibinfo {year} {2020})}\BibitemShut {NoStop}%
\bibitem [{\citenamefont {Zhao}\ \emph {et~al.}(2020)\citenamefont {Zhao}, \citenamefont {Ku}, \citenamefont {Chen}, \citenamefont {Chen}, \citenamefont {Nori}, \citenamefont {Xiang}, \citenamefont {Li}, \citenamefont {Guo},\ and\ \citenamefont {Chen}}]{Zhao2020npj}%
  \BibitemOpen
  \bibfield  {author} {\bibinfo {author} {\bibfnamefont {Y.-Y.}\ \bibnamefont {Zhao}}, \bibinfo {author} {\bibfnamefont {H.-Y.}\ \bibnamefont {Ku}}, \bibinfo {author} {\bibfnamefont {S.-L.}\ \bibnamefont {Chen}}, \bibinfo {author} {\bibfnamefont {H.-B.}\ \bibnamefont {Chen}}, \bibinfo {author} {\bibfnamefont {F.}~\bibnamefont {Nori}}, \bibinfo {author} {\bibfnamefont {G.-Y.}\ \bibnamefont {Xiang}}, \bibinfo {author} {\bibfnamefont {C.-F.}\ \bibnamefont {Li}}, \bibinfo {author} {\bibfnamefont {G.-C.}\ \bibnamefont {Guo}},\ and\ \bibinfo {author} {\bibfnamefont {Y.-N.}\ \bibnamefont {Chen}},\ }\bibfield  {title} {\bibinfo {title} {Experimental demonstration of measurement-device-independent measure of quantum steering},\ }\href {https://doi.org/10.1038/s41534-020-00307-9} {\bibfield  {journal} {\bibinfo  {journal} {npj Quantum Information}\ }\textbf {\bibinfo {volume} {6}},\ \bibinfo {pages} {77} (\bibinfo {year} {2020})}\BibitemShut {NoStop}%
\bibitem [{\citenamefont {{Hsieh}}\ \emph {et~al.}(2023)\citenamefont {{Hsieh}}, \citenamefont {{Ku}},\ and\ \citenamefont {{Budroni}}}]{Hsieh2023arXiv}%
  \BibitemOpen
  \bibfield  {author} {\bibinfo {author} {\bibfnamefont {C.-Y.}\ \bibnamefont {{Hsieh}}}, \bibinfo {author} {\bibfnamefont {H.-Y.}\ \bibnamefont {{Ku}}},\ and\ \bibinfo {author} {\bibfnamefont {C.}~\bibnamefont {{Budroni}}},\ }\bibfield  {title} {\bibinfo {title} {{Characterisation and fundamental limitations of irreversible stochastic steering distillation}},\ }\href {https://doi.org/10.48550/arXiv.2309.06191} {\bibfield  {journal} {\bibinfo  {journal} {arXiv e-prints}\ ,\ \bibinfo {eid} {arXiv:2309.06191}} (\bibinfo {year} {2023})},\ \Eprint {https://arxiv.org/abs/2309.06191} {arXiv:2309.06191} \BibitemShut {NoStop}%
\bibitem [{\citenamefont {Hsieh}\ and\ \citenamefont {Chen}(2024)}]{HsiehPRL2024}%
  \BibitemOpen
  \bibfield  {author} {\bibinfo {author} {\bibfnamefont {C.-Y.}\ \bibnamefont {Hsieh}}\ and\ \bibinfo {author} {\bibfnamefont {S.-L.}\ \bibnamefont {Chen}},\ }\bibfield  {title} {\bibinfo {title} {Thermodynamic approach to quantifying incompatible instruments},\ }\href {https://doi.org/10.1103/PhysRevLett.133.170401} {\bibfield  {journal} {\bibinfo  {journal} {Phys. Rev. Lett.}\ }\textbf {\bibinfo {volume} {133}},\ \bibinfo {pages} {170401} (\bibinfo {year} {2024})}\BibitemShut {NoStop}%
\bibitem [{\citenamefont {{Hsieh}}\ and\ \citenamefont {{Gessner}}(2024)}]{Hsieh2024arXiv}%
  \BibitemOpen
  \bibfield  {author} {\bibinfo {author} {\bibfnamefont {C.-Y.}\ \bibnamefont {{Hsieh}}}\ and\ \bibinfo {author} {\bibfnamefont {M.}~\bibnamefont {{Gessner}}},\ }\bibfield  {title} {\bibinfo {title} {{General quantum resources provide advantages in work extraction tasks}},\ }\href {https://doi.org/10.48550/arXiv.2403.18753} {\bibfield  {journal} {\bibinfo  {journal} {arXiv e-prints}\ ,\ \bibinfo {eid} {arXiv:2403.18753}} (\bibinfo {year} {2024})},\ \Eprint {https://arxiv.org/abs/2403.18753} {arXiv:2403.18753 [quant-ph]} \BibitemShut {NoStop}%
\bibitem [{\citenamefont {Chitambar}\ and\ \citenamefont {Gour}(2019)}]{EricRMP2019}%
  \BibitemOpen
  \bibfield  {author} {\bibinfo {author} {\bibfnamefont {E.}~\bibnamefont {Chitambar}}\ and\ \bibinfo {author} {\bibfnamefont {G.}~\bibnamefont {Gour}},\ }\bibfield  {title} {\bibinfo {title} {Quantum resource theories},\ }\href {https://doi.org/10.1103/RevModPhys.91.025001} {\bibfield  {journal} {\bibinfo  {journal} {Rev. Mod. Phys.}\ }\textbf {\bibinfo {volume} {91}},\ \bibinfo {pages} {025001} (\bibinfo {year} {2019})}\BibitemShut {NoStop}%
\bibitem [{\citenamefont {Hsieh}\ \emph {et~al.}(2025)\citenamefont {Hsieh}, \citenamefont {Stratton}, \citenamefont {Wu},\ and\ \citenamefont {Ku}}]{Hsieh2025PRA}%
  \BibitemOpen
  \bibfield  {author} {\bibinfo {author} {\bibfnamefont {C.-Y.}\ \bibnamefont {Hsieh}}, \bibinfo {author} {\bibfnamefont {B.}~\bibnamefont {Stratton}}, \bibinfo {author} {\bibfnamefont {C.-H.}\ \bibnamefont {Wu}},\ and\ \bibinfo {author} {\bibfnamefont {H.-Y.}\ \bibnamefont {Ku}},\ }\bibfield  {title} {\bibinfo {title} {Dynamical resource theory of incompatibility preservability},\ }\href {https://doi.org/10.1103/PhysRevA.111.022422} {\bibfield  {journal} {\bibinfo  {journal} {Phys. Rev. A}\ }\textbf {\bibinfo {volume} {111}},\ \bibinfo {pages} {022422} (\bibinfo {year} {2025})}\BibitemShut {NoStop}%
\bibitem [{\citenamefont {Moravčíková}\ and\ \citenamefont {Ziman}(2010)}]{Moravcikova_2010}%
  \BibitemOpen
  \bibfield  {author} {\bibinfo {author} {\bibfnamefont {L.}~\bibnamefont {Moravčíková}}\ and\ \bibinfo {author} {\bibfnamefont {M.}~\bibnamefont {Ziman}},\ }\bibfield  {title} {\bibinfo {title} {Entanglement-annihilating and entanglement-breaking channels},\ }\href {https://doi.org/10.1088/1751-8113/43/27/275306} {\bibfield  {journal} {\bibinfo  {journal} {Journal of Physics A: Mathematical and Theoretical}\ }\textbf {\bibinfo {volume} {43}},\ \bibinfo {pages} {275306} (\bibinfo {year} {2010})}\BibitemShut {NoStop}%
\bibitem [{\citenamefont {Hsieh}(2020)}]{Hsieh2020resource}%
  \BibitemOpen
  \bibfield  {author} {\bibinfo {author} {\bibfnamefont {C.-Y.}\ \bibnamefont {Hsieh}},\ }\bibfield  {title} {\bibinfo {title} {Resource {P}reservability},\ }\href {https://doi.org/10.22331/q-2020-03-19-244} {\bibfield  {journal} {\bibinfo  {journal} {{Quantum}}\ }\textbf {\bibinfo {volume} {4}},\ \bibinfo {pages} {244} (\bibinfo {year} {2020})}\BibitemShut {NoStop}%
\bibitem [{\citenamefont {Hsieh}\ \emph {et~al.}(2020)\citenamefont {Hsieh}, \citenamefont {Lostaglio},\ and\ \citenamefont {Ac\'{\i}n}}]{HsiehPRR2020}%
  \BibitemOpen
  \bibfield  {author} {\bibinfo {author} {\bibfnamefont {C.-Y.}\ \bibnamefont {Hsieh}}, \bibinfo {author} {\bibfnamefont {M.}~\bibnamefont {Lostaglio}},\ and\ \bibinfo {author} {\bibfnamefont {A.}~\bibnamefont {Ac\'{\i}n}},\ }\bibfield  {title} {\bibinfo {title} {Entanglement preserving local thermalization},\ }\href {https://doi.org/10.1103/PhysRevResearch.2.013379} {\bibfield  {journal} {\bibinfo  {journal} {Phys. Rev. Res.}\ }\textbf {\bibinfo {volume} {2}},\ \bibinfo {pages} {013379} (\bibinfo {year} {2020})}\BibitemShut {NoStop}%
\bibitem [{\citenamefont {Hsieh}(2021)}]{HsiehPRXQ2021}%
  \BibitemOpen
  \bibfield  {author} {\bibinfo {author} {\bibfnamefont {C.-Y.}\ \bibnamefont {Hsieh}},\ }\bibfield  {title} {\bibinfo {title} {Communication, dynamical resource theory, and thermodynamics},\ }\href {https://doi.org/10.1103/PRXQuantum.2.020318} {\bibfield  {journal} {\bibinfo  {journal} {PRX Quantum}\ }\textbf {\bibinfo {volume} {2}},\ \bibinfo {pages} {020318} (\bibinfo {year} {2021})}\BibitemShut {NoStop}%
\bibitem [{\citenamefont {Takahashi}\ \emph {et~al.}(2022)\citenamefont {Takahashi}, \citenamefont {Rana},\ and\ \citenamefont {Streltsov}}]{TakahashiPRA2022}%
  \BibitemOpen
  \bibfield  {author} {\bibinfo {author} {\bibfnamefont {M.}~\bibnamefont {Takahashi}}, \bibinfo {author} {\bibfnamefont {S.}~\bibnamefont {Rana}},\ and\ \bibinfo {author} {\bibfnamefont {A.}~\bibnamefont {Streltsov}},\ }\bibfield  {title} {\bibinfo {title} {Creating and destroying coherence with quantum channels},\ }\href {https://doi.org/10.1103/PhysRevA.105.L060401} {\bibfield  {journal} {\bibinfo  {journal} {Phys. Rev. A}\ }\textbf {\bibinfo {volume} {105}},\ \bibinfo {pages} {L060401} (\bibinfo {year} {2022})}\BibitemShut {NoStop}%
\bibitem [{\citenamefont {Stratton}\ \emph {et~al.}(2024)\citenamefont {Stratton}, \citenamefont {Hsieh},\ and\ \citenamefont {Skrzypczyk}}]{StrattonPRL2024}%
  \BibitemOpen
  \bibfield  {author} {\bibinfo {author} {\bibfnamefont {B.}~\bibnamefont {Stratton}}, \bibinfo {author} {\bibfnamefont {C.-Y.}\ \bibnamefont {Hsieh}},\ and\ \bibinfo {author} {\bibfnamefont {P.}~\bibnamefont {Skrzypczyk}},\ }\bibfield  {title} {\bibinfo {title} {Dynamical resource theory of informational nonequilibrium preservability},\ }\href {https://doi.org/10.1103/PhysRevLett.132.110202} {\bibfield  {journal} {\bibinfo  {journal} {Phys. Rev. Lett.}\ }\textbf {\bibinfo {volume} {132}},\ \bibinfo {pages} {110202} (\bibinfo {year} {2024})}\BibitemShut {NoStop}%
\bibitem [{\citenamefont {Popescu}(1995)}]{SanduPRL1995}%
  \BibitemOpen
  \bibfield  {author} {\bibinfo {author} {\bibfnamefont {S.}~\bibnamefont {Popescu}},\ }\bibfield  {title} {\bibinfo {title} {Bell's inequalities and density matrices: Revealing ``hidden'' nonlocality},\ }\href {https://doi.org/10.1103/PhysRevLett.74.2619} {\bibfield  {journal} {\bibinfo  {journal} {Phys. Rev. Lett.}\ }\textbf {\bibinfo {volume} {74}},\ \bibinfo {pages} {2619} (\bibinfo {year} {1995})}\BibitemShut {NoStop}%
\bibitem [{\citenamefont {Hirsch}\ \emph {et~al.}(2013)\citenamefont {Hirsch}, \citenamefont {Quintino}, \citenamefont {Bowles},\ and\ \citenamefont {Brunner}}]{HirschPRL2013}%
  \BibitemOpen
  \bibfield  {author} {\bibinfo {author} {\bibfnamefont {F.}~\bibnamefont {Hirsch}}, \bibinfo {author} {\bibfnamefont {M.~T.}\ \bibnamefont {Quintino}}, \bibinfo {author} {\bibfnamefont {J.}~\bibnamefont {Bowles}},\ and\ \bibinfo {author} {\bibfnamefont {N.}~\bibnamefont {Brunner}},\ }\bibfield  {title} {\bibinfo {title} {Genuine hidden quantum nonlocality},\ }\href {https://doi.org/10.1103/PhysRevLett.111.160402} {\bibfield  {journal} {\bibinfo  {journal} {Phys. Rev. Lett.}\ }\textbf {\bibinfo {volume} {111}},\ \bibinfo {pages} {160402} (\bibinfo {year} {2013})}\BibitemShut {NoStop}%
\bibitem [{\citenamefont {Quintino}\ \emph {et~al.}(2016)\citenamefont {Quintino}, \citenamefont {Brunner},\ and\ \citenamefont {Huber}}]{QuintinohysPRA2016}%
  \BibitemOpen
  \bibfield  {author} {\bibinfo {author} {\bibfnamefont {M.~T.}\ \bibnamefont {Quintino}}, \bibinfo {author} {\bibfnamefont {N.}~\bibnamefont {Brunner}},\ and\ \bibinfo {author} {\bibfnamefont {M.}~\bibnamefont {Huber}},\ }\bibfield  {title} {\bibinfo {title} {Superactivation of quantum steering},\ }\href {https://doi.org/10.1103/PhysRevA.94.062123} {\bibfield  {journal} {\bibinfo  {journal} {Phys. Rev. A}\ }\textbf {\bibinfo {volume} {94}},\ \bibinfo {pages} {062123} (\bibinfo {year} {2016})}\BibitemShut {NoStop}%
\bibitem [{\citenamefont {Ku}\ \emph {et~al.}(2021)\citenamefont {Ku}, \citenamefont {Weng}, \citenamefont {Shih}, \citenamefont {Kuo}, \citenamefont {Lambert}, \citenamefont {Nori}, \citenamefont {Chuu},\ and\ \citenamefont {Chen}}]{KuPRR2021}%
  \BibitemOpen
  \bibfield  {author} {\bibinfo {author} {\bibfnamefont {H.-Y.}\ \bibnamefont {Ku}}, \bibinfo {author} {\bibfnamefont {H.-C.}\ \bibnamefont {Weng}}, \bibinfo {author} {\bibfnamefont {Y.-A.}\ \bibnamefont {Shih}}, \bibinfo {author} {\bibfnamefont {P.-C.}\ \bibnamefont {Kuo}}, \bibinfo {author} {\bibfnamefont {N.}~\bibnamefont {Lambert}}, \bibinfo {author} {\bibfnamefont {F.}~\bibnamefont {Nori}}, \bibinfo {author} {\bibfnamefont {C.-S.}\ \bibnamefont {Chuu}},\ and\ \bibinfo {author} {\bibfnamefont {Y.-N.}\ \bibnamefont {Chen}},\ }\bibfield  {title} {\bibinfo {title} {Hidden nonmacrorealism: Reviving the leggett-garg inequality with stochastic operations},\ }\href {https://doi.org/10.1103/PhysRevResearch.3.043083} {\bibfield  {journal} {\bibinfo  {journal} {Phys. Rev. Res.}\ }\textbf {\bibinfo {volume} {3}},\ \bibinfo {pages} {043083} (\bibinfo {year} {2021})}\BibitemShut {NoStop}%
\bibitem [{\citenamefont {Ku}\ \emph {et~al.}(2023)\citenamefont {Ku}, \citenamefont {Lee}, \citenamefont {Lai}, \citenamefont {Lin},\ and\ \citenamefont {Chen}}]{KuPRA2023}%
  \BibitemOpen
  \bibfield  {author} {\bibinfo {author} {\bibfnamefont {H.-Y.}\ \bibnamefont {Ku}}, \bibinfo {author} {\bibfnamefont {K.-Y.}\ \bibnamefont {Lee}}, \bibinfo {author} {\bibfnamefont {P.-R.}\ \bibnamefont {Lai}}, \bibinfo {author} {\bibfnamefont {J.-D.}\ \bibnamefont {Lin}},\ and\ \bibinfo {author} {\bibfnamefont {Y.-N.}\ \bibnamefont {Chen}},\ }\bibfield  {title} {\bibinfo {title} {Coherent activation of a steerability-breaking channel},\ }\href {https://doi.org/10.1103/PhysRevA.107.042415} {\bibfield  {journal} {\bibinfo  {journal} {Phys. Rev. A}\ }\textbf {\bibinfo {volume} {107}},\ \bibinfo {pages} {042415} (\bibinfo {year} {2023})}\BibitemShut {NoStop}%
\bibitem [{\citenamefont {Villegas-Aguilar}\ \emph {et~al.}(2024)\citenamefont {Villegas-Aguilar}, \citenamefont {Polino}, \citenamefont {Ghafari}, \citenamefont {Quintino}, \citenamefont {Laverick}, \citenamefont {Berkman}, \citenamefont {Rogge}, \citenamefont {Shalm}, \citenamefont {Tischler}, \citenamefont {Cavalcanti}, \citenamefont {Slussarenko},\ and\ \citenamefont {Pryde}}]{VillegasAguilar2024}%
  \BibitemOpen
  \bibfield  {author} {\bibinfo {author} {\bibfnamefont {L.}~\bibnamefont {Villegas-Aguilar}}, \bibinfo {author} {\bibfnamefont {E.}~\bibnamefont {Polino}}, \bibinfo {author} {\bibfnamefont {F.}~\bibnamefont {Ghafari}}, \bibinfo {author} {\bibfnamefont {M.~T.}\ \bibnamefont {Quintino}}, \bibinfo {author} {\bibfnamefont {K.~T.}\ \bibnamefont {Laverick}}, \bibinfo {author} {\bibfnamefont {I.~R.}\ \bibnamefont {Berkman}}, \bibinfo {author} {\bibfnamefont {S.}~\bibnamefont {Rogge}}, \bibinfo {author} {\bibfnamefont {L.~K.}\ \bibnamefont {Shalm}}, \bibinfo {author} {\bibfnamefont {N.}~\bibnamefont {Tischler}}, \bibinfo {author} {\bibfnamefont {E.~G.}\ \bibnamefont {Cavalcanti}}, \bibinfo {author} {\bibfnamefont {S.}~\bibnamefont {Slussarenko}},\ and\ \bibinfo {author} {\bibfnamefont {G.~J.}\ \bibnamefont {Pryde}},\ }\bibfield  {title} {\bibinfo {title} {Nonlocality activation in a photonic quantum network},\ }\bibfield  {journal} {\bibinfo  {journal} {Nature Communications}\ }\textbf {\bibinfo {volume} {15}},\
  \href {https://doi.org/10.1038/s41467-024-47354-w} {10.1038/s41467-024-47354-w} (\bibinfo {year} {2024})\BibitemShut {NoStop}%
\bibitem [{\citenamefont {{Tabia}}\ \emph {et~al.}(2024)\citenamefont {{Tabia}}, \citenamefont {{Hsieh}},\ and\ \citenamefont {{Hsieh}}}]{Gelo2024arXiv}%
  \BibitemOpen
  \bibfield  {author} {\bibinfo {author} {\bibfnamefont {G.~N.~M.}\ \bibnamefont {{Tabia}}}, \bibinfo {author} {\bibfnamefont {C.-Y.}\ \bibnamefont {{Hsieh}}},\ and\ \bibinfo {author} {\bibfnamefont {M.-H.}\ \bibnamefont {{Hsieh}}},\ }\bibfield  {title} {\bibinfo {title} {{Bell nonlocality from compatibility of entanglement-breaking channels}},\ }\href {https://doi.org/10.48550/arXiv.2410.13499} {\bibfield  {journal} {\bibinfo  {journal} {arXiv e-prints}\ ,\ \bibinfo {eid} {arXiv:2410.13499}} (\bibinfo {year} {2024})},\ \Eprint {https://arxiv.org/abs/2410.13499} {arXiv:2410.13499 [quant-ph]} \BibitemShut {NoStop}%
\bibitem [{\citenamefont {{Arai}}\ and\ \citenamefont {{Hayashi}}(2024)}]{Arai2024arXiv}%
  \BibitemOpen
  \bibfield  {author} {\bibinfo {author} {\bibfnamefont {H.}~\bibnamefont {{Arai}}}\ and\ \bibinfo {author} {\bibfnamefont {M.}~\bibnamefont {{Hayashi}}},\ }\bibfield  {title} {\bibinfo {title} {{Quantum Implementation of Non-Positive-Operator-Valued Measurements in General Probabilistic Theories by Post-Selected POVMs}},\ }\href {https://doi.org/10.48550/arXiv.2411.01873} {\bibfield  {journal} {\bibinfo  {journal} {arXiv e-prints}\ ,\ \bibinfo {eid} {arXiv:2411.01873}} (\bibinfo {year} {2024})},\ \Eprint {https://arxiv.org/abs/2411.01873} {arXiv:2411.01873 [quant-ph]} \BibitemShut {NoStop}%
\bibitem [{\citenamefont {Ku}\ \emph {et~al.}(2026)\citenamefont {Ku}, \citenamefont {Hsieh},\ and\ \citenamefont {Budroni}}]{Ku2026npj}%
  \BibitemOpen
  \bibfield  {author} {\bibinfo {author} {\bibfnamefont {H.-Y.}\ \bibnamefont {Ku}}, \bibinfo {author} {\bibfnamefont {C.-Y.}\ \bibnamefont {Hsieh}},\ and\ \bibinfo {author} {\bibfnamefont {C.}~\bibnamefont {Budroni}},\ }\bibfield  {title} {\bibinfo {title} {No-go theorems on probabilistically enhancing measurement incompatibility’s advantages},\ }\bibfield  {journal} {\bibinfo  {journal} {npj Quantum Information}\ }\href {https://doi.org/10.1038/s41534-026-01270-7} {10.1038/s41534-026-01270-7} (\bibinfo {year} {2026})\BibitemShut {NoStop}%
\bibitem [{\citenamefont {Wiseman}\ \emph {et~al.}(2007)\citenamefont {Wiseman}, \citenamefont {Jones},\ and\ \citenamefont {Doherty}}]{WisemanPRL2007}%
  \BibitemOpen
  \bibfield  {author} {\bibinfo {author} {\bibfnamefont {H.~M.}\ \bibnamefont {Wiseman}}, \bibinfo {author} {\bibfnamefont {S.~J.}\ \bibnamefont {Jones}},\ and\ \bibinfo {author} {\bibfnamefont {A.~C.}\ \bibnamefont {Doherty}},\ }\bibfield  {title} {\bibinfo {title} {Steering, entanglement, nonlocality, and the einstein-podolsky-rosen paradox},\ }\href {https://doi.org/10.1103/PhysRevLett.98.140402} {\bibfield  {journal} {\bibinfo  {journal} {Phys. Rev. Lett.}\ }\textbf {\bibinfo {volume} {98}},\ \bibinfo {pages} {140402} (\bibinfo {year} {2007})}\BibitemShut {NoStop}%
\bibitem [{\citenamefont {Quintino}\ \emph {et~al.}(2014{\natexlab{b}})\citenamefont {Quintino}, \citenamefont {V\'ertesi},\ and\ \citenamefont {Brunner}}]{Quantino2014PRL}%
  \BibitemOpen
  \bibfield  {author} {\bibinfo {author} {\bibfnamefont {M.~T.}\ \bibnamefont {Quintino}}, \bibinfo {author} {\bibfnamefont {T.}~\bibnamefont {V\'ertesi}},\ and\ \bibinfo {author} {\bibfnamefont {N.}~\bibnamefont {Brunner}},\ }\bibfield  {title} {\bibinfo {title} {Joint measurability, einstein-podolsky-rosen steering, and bell nonlocality},\ }\href {https://doi.org/10.1103/PhysRevLett.113.160402} {\bibfield  {journal} {\bibinfo  {journal} {Phys. Rev. Lett.}\ }\textbf {\bibinfo {volume} {113}},\ \bibinfo {pages} {160402} (\bibinfo {year} {2014}{\natexlab{b}})}\BibitemShut {NoStop}%
\bibitem [{\citenamefont {Uola}\ \emph {et~al.}(2014)\citenamefont {Uola}, \citenamefont {Moroder},\ and\ \citenamefont {G\"uhne}}]{Uola2014PRL}%
  \BibitemOpen
  \bibfield  {author} {\bibinfo {author} {\bibfnamefont {R.}~\bibnamefont {Uola}}, \bibinfo {author} {\bibfnamefont {T.}~\bibnamefont {Moroder}},\ and\ \bibinfo {author} {\bibfnamefont {O.}~\bibnamefont {G\"uhne}},\ }\bibfield  {title} {\bibinfo {title} {Joint measurability of generalized measurements implies classicality},\ }\href {https://doi.org/10.1103/PhysRevLett.113.160403} {\bibfield  {journal} {\bibinfo  {journal} {Phys. Rev. Lett.}\ }\textbf {\bibinfo {volume} {113}},\ \bibinfo {pages} {160403} (\bibinfo {year} {2014})}\BibitemShut {NoStop}%
\bibitem [{\citenamefont {Uola}\ \emph {et~al.}(2015)\citenamefont {Uola}, \citenamefont {Budroni}, \citenamefont {G\"uhne},\ and\ \citenamefont {Pellonp\"a\"a}}]{Uola2015PRL}%
  \BibitemOpen
  \bibfield  {author} {\bibinfo {author} {\bibfnamefont {R.}~\bibnamefont {Uola}}, \bibinfo {author} {\bibfnamefont {C.}~\bibnamefont {Budroni}}, \bibinfo {author} {\bibfnamefont {O.}~\bibnamefont {G\"uhne}},\ and\ \bibinfo {author} {\bibfnamefont {J.-P.}\ \bibnamefont {Pellonp\"a\"a}},\ }\bibfield  {title} {\bibinfo {title} {One-to-one mapping between steering and joint measurability problems},\ }\href {https://doi.org/10.1103/PhysRevLett.115.230402} {\bibfield  {journal} {\bibinfo  {journal} {Phys. Rev. Lett.}\ }\textbf {\bibinfo {volume} {115}},\ \bibinfo {pages} {230402} (\bibinfo {year} {2015})}\BibitemShut {NoStop}%
\bibitem [{\citenamefont {Ku}\ \emph {et~al.}(2022{\natexlab{a}})\citenamefont {Ku}, \citenamefont {Kadlec}, \citenamefont {\ifmmode~\check{C}\else \v{C}\fi{}ernoch}, \citenamefont {Quintino}, \citenamefont {Zhou}, \citenamefont {Lemr}, \citenamefont {Lambert}, \citenamefont {Miranowicz}, \citenamefont {Chen}, \citenamefont {Nori},\ and\ \citenamefont {Chen}}]{KuPRXQ2022}%
  \BibitemOpen
  \bibfield  {author} {\bibinfo {author} {\bibfnamefont {H.-Y.}\ \bibnamefont {Ku}}, \bibinfo {author} {\bibfnamefont {J.}~\bibnamefont {Kadlec}}, \bibinfo {author} {\bibfnamefont {A.}~\bibnamefont {\ifmmode~\check{C}\else \v{C}\fi{}ernoch}}, \bibinfo {author} {\bibfnamefont {M.~T.}\ \bibnamefont {Quintino}}, \bibinfo {author} {\bibfnamefont {W.}~\bibnamefont {Zhou}}, \bibinfo {author} {\bibfnamefont {K.}~\bibnamefont {Lemr}}, \bibinfo {author} {\bibfnamefont {N.}~\bibnamefont {Lambert}}, \bibinfo {author} {\bibfnamefont {A.}~\bibnamefont {Miranowicz}}, \bibinfo {author} {\bibfnamefont {S.-L.}\ \bibnamefont {Chen}}, \bibinfo {author} {\bibfnamefont {F.}~\bibnamefont {Nori}},\ and\ \bibinfo {author} {\bibfnamefont {Y.-N.}\ \bibnamefont {Chen}},\ }\bibfield  {title} {\bibinfo {title} {Quantifying quantumness of channels without entanglement},\ }\href {https://doi.org/10.1103/PRXQuantum.3.020338} {\bibfield  {journal} {\bibinfo  {journal} {PRX Quantum}\ }\textbf {\bibinfo {volume} {3}},\ \bibinfo {pages}
  {020338} (\bibinfo {year} {2022}{\natexlab{a}})}\BibitemShut {NoStop}%
\bibitem [{\citenamefont {Ku}\ \emph {et~al.}(2022{\natexlab{b}})\citenamefont {Ku}, \citenamefont {Hsieh}, \citenamefont {Chen}, \citenamefont {Chen},\ and\ \citenamefont {Budroni}}]{Ku2022NC}%
  \BibitemOpen
  \bibfield  {author} {\bibinfo {author} {\bibfnamefont {H.-Y.}\ \bibnamefont {Ku}}, \bibinfo {author} {\bibfnamefont {C.-Y.}\ \bibnamefont {Hsieh}}, \bibinfo {author} {\bibfnamefont {S.-L.}\ \bibnamefont {Chen}}, \bibinfo {author} {\bibfnamefont {Y.-N.}\ \bibnamefont {Chen}},\ and\ \bibinfo {author} {\bibfnamefont {C.}~\bibnamefont {Budroni}},\ }\bibfield  {title} {\bibinfo {title} {Complete classification of steerability under local filters and its relation with measurement incompatibility},\ }\href {https://doi.org/10.1038/s41467-022-32466-y} {\bibfield  {journal} {\bibinfo  {journal} {Nat. Commun.}\ }\textbf {\bibinfo {volume} {13}},\ \bibinfo {pages} {4973} (\bibinfo {year} {2022}{\natexlab{b}})}\BibitemShut {NoStop}%
\bibitem [{\citenamefont {Wang}\ \emph {et~al.}(2024)\citenamefont {Wang}, \citenamefont {Ku}, \citenamefont {Lin},\ and\ \citenamefont {Chen}}]{Wang2024}%
  \BibitemOpen
  \bibfield  {author} {\bibinfo {author} {\bibfnamefont {H.-M.}\ \bibnamefont {Wang}}, \bibinfo {author} {\bibfnamefont {H.-Y.}\ \bibnamefont {Ku}}, \bibinfo {author} {\bibfnamefont {J.-Y.}\ \bibnamefont {Lin}},\ and\ \bibinfo {author} {\bibfnamefont {H.-B.}\ \bibnamefont {Chen}},\ }\bibfield  {title} {\bibinfo {title} {Deep learning the hierarchy of steering measurement settings of qubit-pair states},\ }\href {https://doi.org/10.1038/s42005-024-01563-3} {\bibfield  {journal} {\bibinfo  {journal} {Communications Physics}\ }\textbf {\bibinfo {volume} {7}},\ \bibinfo {pages} {72} (\bibinfo {year} {2024})}\BibitemShut {NoStop}%
\bibitem [{\citenamefont {Chang}\ \emph {et~al.}(2024)\citenamefont {Chang}, \citenamefont {Ju}, \citenamefont {Chen}, \citenamefont {Chen},\ and\ \citenamefont {Ku}}]{ChangPRR2024}%
  \BibitemOpen
  \bibfield  {author} {\bibinfo {author} {\bibfnamefont {W.-G.}\ \bibnamefont {Chang}}, \bibinfo {author} {\bibfnamefont {C.-Y.}\ \bibnamefont {Ju}}, \bibinfo {author} {\bibfnamefont {G.-Y.}\ \bibnamefont {Chen}}, \bibinfo {author} {\bibfnamefont {Y.-N.}\ \bibnamefont {Chen}},\ and\ \bibinfo {author} {\bibfnamefont {H.-Y.}\ \bibnamefont {Ku}},\ }\bibfield  {title} {\bibinfo {title} {Visually quantifying single-qubit quantum memory},\ }\href {https://doi.org/10.1103/PhysRevResearch.6.023035} {\bibfield  {journal} {\bibinfo  {journal} {Phys. Rev. Res.}\ }\textbf {\bibinfo {volume} {6}},\ \bibinfo {pages} {023035} (\bibinfo {year} {2024})}\BibitemShut {NoStop}%
\bibitem [{\citenamefont {Bavaresco}\ \emph {et~al.}(2017)\citenamefont {Bavaresco}, \citenamefont {Quintino}, \citenamefont {Guerini}, \citenamefont {Maciel}, \citenamefont {Cavalcanti},\ and\ \citenamefont {Cunha}}]{JessicaPRA2017}%
  \BibitemOpen
  \bibfield  {author} {\bibinfo {author} {\bibfnamefont {J.}~\bibnamefont {Bavaresco}}, \bibinfo {author} {\bibfnamefont {M.~T.}\ \bibnamefont {Quintino}}, \bibinfo {author} {\bibfnamefont {L.}~\bibnamefont {Guerini}}, \bibinfo {author} {\bibfnamefont {T.~O.}\ \bibnamefont {Maciel}}, \bibinfo {author} {\bibfnamefont {D.}~\bibnamefont {Cavalcanti}},\ and\ \bibinfo {author} {\bibfnamefont {M.~T.}\ \bibnamefont {Cunha}},\ }\bibfield  {title} {\bibinfo {title} {Most incompatible measurements for robust steering tests},\ }\href {https://doi.org/10.1103/PhysRevA.96.022110} {\bibfield  {journal} {\bibinfo  {journal} {Phys. Rev. A}\ }\textbf {\bibinfo {volume} {96}},\ \bibinfo {pages} {022110} (\bibinfo {year} {2017})}\BibitemShut {NoStop}%
\end{thebibliography}

\providecommand{\noopsort}[1]{}\providecommand{\singleletter}[1]{#1}%

\appendix
\section{Derivation of dual SDP}
    \label{appendix:dual problem derivation}
    In this section, we show how to obtain the dual SDP in Eq.~\eqref{eq:optimization JM SDP dual}. Recall the primal problem in Eq.~\eqref{eq:optimization JM SDP}. For convenience, we introduce an auxiliary scalar variable $\mu$ through the normalization condition \(\sum_\lambda G_\lambda=\mu\openone\), and rewrite the primal problem as below:
    \begin{equation}
        \begin{aligned}
            \text{Given } &\mathcal{N}, \mathbf{M} \\
            \text{Find } &\min_{\mu, t, W, \{G_\lambda\}_{\lambda}} \mu
            - 1 = \mathcal{R}^{JM}_{IP}(\mathcal{N}; \mathbf{M})\\
            \text{s.t. } 
            &\mathcal{N}^\dagger(M_{a|x}) 
            + \operatorname{Tr}_{out}[
            (M_{a|x}^T \otimes \openone_{in})W] \\
            &= \sum_\lambda D(a|x, \lambda)G_\lambda,
            \quad \forall a,x \\
            &\operatorname{Tr}_{out}(W) = t\openone_{in},
            \quad W \geq 0 \\
            &\sum_\lambda G_\lambda = \mu \openone \\
            &G_\lambda \geq 0, \quad \forall\lambda \\
            &t \geq 0 .
            \label{eq:rewrite primal}
        \end{aligned}
    \end{equation}
    By the first constraint
    \begin{equation}
        \begin{aligned}
            &\mathcal{N}^\dagger(M_{a|x}) + \operatorname{Tr}_{out}[(M_{a|x}^T \otimes \openone_{in})W] \\&= \sum_\lambda D(a|x, \lambda)G_\lambda, \quad \forall a,x
            \label{eq:first equality constraint}
        \end{aligned}
    \end{equation}
    $\mu$ must satisfy \(\mu=1+t\). We can then define a series of dual variables \(Y_{a|x} = Y_{a|x}^\dagger\), \(S = S^\dagger\), and \(Z = Z^\dagger\) such that the following Lagrangian can be obtained:
    \begin{equation}
        \begin{aligned}
            L = &(\mu - 1) \\
            &+ \sum_{a, x} \operatorname{Tr} \Biggl[ Y_{a|x} \biggl(\mathcal{N}^\dagger(M_{a|x}) + \operatorname{Tr}_{out} \Bigl[(M_{a|x}^T \otimes \openone_{in} )W \Bigr] \\
            & \qquad \qquad \quad - \sum_\lambda D(a|x, \lambda) G_\lambda \biggr) \Biggr] \\
            &+ \operatorname{Tr}\Bigl[ S \bigl(t\openone_{in} - \operatorname{Tr}_{out}(W) \bigr) \Bigr] \\
            &+ \operatorname{Tr} \left[ Z \left( \sum_\lambda G_\lambda - \mu\openone \right) \right]
        \end{aligned}
        \label{eq:SDP Lagrangian}
    \end{equation}
    We can reformulate the Lagrangian and have:
    \begin{equation}
        \begin{aligned}
            L = &\sum_{a, x} \operatorname{Tr}[Y_{a|x} \mathcal{N}^\dagger(M_{a|x})]-1 \\
            &+ \mu(1-\operatorname{Tr}[Z]) + t\operatorname{Tr}[S] \\
            &+ \operatorname{Tr}\left[\left(\sum_{a, x} M_{a|x}^T \otimes Y_{a|x} - \openone \otimes S\right)W\right] \\
            &+ \sum_\lambda \operatorname{Tr} \left[\left(z-\sum_{a, x} D(a|x, \lambda)Y_{a|x}\right)G_\lambda\right].
        \end{aligned}
        \label{eq:SDP Lagrangian organized}
    \end{equation}
    To write it down, note that the Lagrangian is independent of the primal variables when \begin{equation}
        \begin{aligned}
            &1 - \operatorname{Tr}[Z] = 0, \\
        &\operatorname{Tr}[S] = 0.
        &\sum_{a, x} M_{a|x}^T \otimes Y_{a|x} - \openone \otimes S = 0.
        \end{aligned} 
        \label{eq:dual constraint 1}
    \end{equation}
    Finally, we can relax the above equalities to inequalities such that 
   the dual problem can be derived as
    \begin{equation*}
        \begin{aligned}
            \text{Given } &\mathcal{N}, \mathbf{M}\\
            \text{Find } &\max_{\mathbf{Y}, Z, S} \sum_{a, x}\operatorname{Tr}[Y_{a|x}\mathcal{N}^\dagger(M_{a|x})] - 1 \\
            &=\mathcal{R}^{JM}_{IP}(\mathcal{N}; \mathbf{M})\\
            \text{s.t. } & Y_{a|x} = Y^\dagger_{a|x}, \quad \forall a, x\\
            &\sum_{a, x} M^T_{a|x} \otimes Y_{a|x} \geq \openone \otimes S \\
            &\operatorname{Tr}[S] \geq 0, \quad S = S^\dagger\\
            &Z \geq \sum_{a, x} D(a|x, \lambda)Y_{a|x} \\
            &\operatorname{Tr}[Z] \leq 1, Z = Z^\dagger.
        \end{aligned}
    \end{equation*}

\section{Proof of Result~\ref{result:monotone}}
    \label{appendix:monotone proof}
    In this section, we will prove that $\mathcal{R}_{IP}^{JM}$ satisfy the resource monotones, namely faithfulness, monotonicity under allowed operations, and convexity.
    
    \paragraph{Faithfulness.}
    By definition, an incompatibility-annihilating channel maps every measurement assemblage to a jointly measurable one. Hence, , by definition, we always have the following for any $\mathcal{N} \in \mathbb{IA}$
    \[
        \mathcal{R}_{IP}^{JM}(\mathcal{N})=0,
        \qquad \forall \mathcal{N}\in\mathbb{IA}.
    \]
    
    \paragraph{Monotonicity under allowed operations.}
    We now prove monotonicity under allowed operations. Consider
    \(\Lambda\in\mathbb{AO}\) as an allowed operation, and the given measurement assemblage \(\mathbf{M}=\{M_{a|x}\}_{a,x}\). Let
    \(t^*\) and \(\mathcal{W}^*\) be optimal solutions for
    \(\mathcal{R}_{IP}^{JM}(\mathcal{N};\mathbf{M})\).The mixing channel
    \[
        \mathcal{N}_{t^*}
        \coloneqq
        \frac{\mathcal{N}+t^*\mathcal{W}^*}{1+t^*}
    \]
    must generate a jointly-measurable measurement assemblage:
    \[
        \mathcal{N}_{t^*}(M_{a|x})=M^{JM}_{a|x},
        \qquad \forall a,x.
    \]
    %for some jointly measurable measurement assemblage\(\{M^{JM}_{a|x}\}_{a,x}\).
    
    Since \(\Lambda\) is an allowed operation, it maps a free resource to another free resource. This fact preserves the relevant joint-measurability condition. Moreover,
    by linearity of \(\Lambda\), we have
    \[
        \Lambda(\mathcal{N}_{t^*})
        =
        \frac{\Lambda(\mathcal{N})+t^*\Lambda(\mathcal{W}^*)}{1+t^*}.
    \]
    Therefore, the pair \((t^*,\Lambda(\mathcal{W}^*))\) gives a feasible solution
    for the robustness optimization of \(\Lambda(\mathcal{N})\). Hence,
    \[
        \mathcal{R}_{IP}^{JM}(\Lambda(\mathcal{N});\mathbf{M})
        \leq
        \mathcal{R}_{IP}^{JM}(\mathcal{N};\mathbf{M}).
    \]
    Taking the maximum over all admissible measurement assemblages gives
    \[
        \mathcal{R}_{IP}^{JM}(\Lambda(\mathcal{N}))
        \leq
        \mathcal{R}_{IP}^{JM}(\mathcal{N}),
        \qquad \forall \Lambda\in\mathbb{AO}.
    \]
    %We start from prove the monotonicity of $\mathcal{R}_{IP, \space max}^{JM}$. First, fix a measurement assemblage $\mathbf{M}$. Let $t^*$ and $\mathcal{W}^*$ be optimal solutions of $\mathcal{R}_{IP, \space max}^{JM}(\mathcal{N};\mathbf{M})$ for the channel $\mathcal{N}$, i.e.,
    %\begin{equation*}
        %\mathcal{N}_{t^*} \coloneqq
        %\frac{\mathcal{N}+t^*\mathcal{W}^*}{1+t^*} \in \mathbb{IA}.
    %\end{equation*}
    %Since $\Lambda \in \mathbb{AO}$ is an allowed operation, it preserves the set $\mathbb{IA}$. Therefore,
    %\begin{equation*}
        %\Lambda(\mathcal{N}_{t^*})
        %= \frac{\Lambda(\mathcal{N}) + t^*\Lambda(\mathcal{W}^*)}{1+t^*}
        %\in \mathbb{IA}.
    %\end{equation*}
    %This shows that $(t^*,\,\Lambda(\mathcal{W}^*))$ is a feasible solution of Eq.~\eqref{eq:optimization JM} for the channel $\Lambda(\mathcal{N})$. Consequently, the optimal value for $\Lambda(\mathcal{N})$ cannot exceed $t^*$, and hence
    %\begin{equation*}
        %\mathcal{R}_{IP}^{JM}(\Lambda(\mathcal{N}); \mathbf{M}) \leq \mathcal{R}_{IP}^{JM}(\mathcal{N}; \mathbf{M}),
        %\quad \forall \Lambda \in \mathbb{AO}.
    %\end{equation*}
    %Since the above argument holds for any fixed measurement assemblage
    %$\mathbf{M}$, taking the maximum over $\mathbf{M}$ completes the proof.

    %The proof of monotonicity for $\mathcal{R}_{IP, \space max}^{LHS}$ follows analogously by noting that allowed operations preserve the set of LHS assemblages.

    \paragraph{Convexity.}
    %We next prove convexity. 
    Given a measurement assemblage \(\mathbf{M}=\{M_{a|x}\}_{a,x}\), probability $p(i)$, convex mixture of channel
    \(
        \mathcal{N}
        \coloneqq
        \sum_i p(i)\mathcal{N}_i .
    \) as well as the optimal solutions of
    \(\mathcal{R}_{IP}^{JM}(\mathcal{N}_i;\mathbf{M})\) as \(t_i^*\) and
    \(\mathcal{W}_i^*\) for each \(i\). We can define
    \[
        \mathcal{N}_{i,t_i^*}
        \coloneqq
        \frac{\mathcal{N}_i+t_i^*\mathcal{W}_i^*}{1+t_i^*}.
    \]
    By the definition of robustness, we always have
    \[
        \mathcal{N}_{i,t_i^*}(M_{a|x})=M^{JM}_{i,a|x},
        \qquad \forall a,x,
    \]
    where \(\{M^{JM}_{i,a|x}\}_{a,x}\) is jointly measurable for each \(i\).

    We can further define the following 
    \[
        t\coloneqq \sum_i p(i)t_i^*,
    \]
    and, a valid channel
    \[
        \mathcal{W}
        \coloneqq
        \frac{1}{t}\sum_i p(i)t_i^*\mathcal{W}_i^*.
    \]
    Note that in the trivial case \(t=0\), all $t_i^*$ must also be 0, and by the convexity of the free resource, $\sum_i p(i)N_i$ is IA. Thus, we only focus on $t_i>0$ for some $i$.
    %Since the set of channels \(\mathbb{C}\) is convex, we have
    %\(\mathcal{W}\in\mathbb{C}\).

   The direct calculation gives us
    \[
    \begin{aligned}
        \frac{\mathcal{N}+t\mathcal{W}}{1+t}
        &=
        \frac{1}{1+t}
        \left(
            \sum_i p(i)\mathcal{N}_i
            +
            \sum_i p(i)t_i^*\mathcal{W}_i^*
        \right) \\
        &=
        \sum_i
        \frac{p(i)(1+t_i^*)}{1+t}
        \frac{\mathcal{N}_i+t_i^*\mathcal{W}_i^*}{1+t_i^*} \\
        &=
        \sum_i q_i \mathcal{N}_{i,t_i^*},
    \end{aligned}
    \]
    where
    \[
        q_i
        \coloneqq
        \frac{p(i)(1+t_i^*)}{1+t}.
    \]
    Clearly \(q_i\geq 0\) and
    \[
        \sum_i q_i
        =
        \frac{\sum_i p(i)(1+t_i^*)}{1+t}
        =
        1.
    \]
    Therefore,
    \[
        \frac{\mathcal{N}+t\mathcal{W}}{1+t}
    \]
    is a valid mixture of channels. Moreover, it always produces
    jointly measurable assemblages for any input  \(\mathbf{M}\). %Since the set of jointly measurable
    %measurement assemblages is convex, the resulting measurement assemblage is
    %also jointly measurable. 
    Hence \(t\) is a feasible solution for
    \(\mathcal{R}_{IP}^{JM}(\sum_i p(i)\mathcal{N}_i;\mathbf{M})\) but may not be necessarily optimal, and thus, we obtain
    \[
        \begin{aligned}
            \mathcal{R}_{IP}^{JM}
            \left(
                \sum_i p(i)\mathcal{N}_i;\mathbf{M}
            \right)
            &\leq
            \sum_i p(i)t_i^*
            \\&=
            \sum_i p(i)
            \mathcal{R}_{IP}^{JM}(\mathcal{N}_i;\mathbf{M}).
        \end{aligned}
    \]
    Finally, taking the maximum over \(\mathbf{M}\), we get
    \[
    \begin{aligned}
        \mathcal{R}_{IP}^{JM}
        \left(
            \sum_i p(i)\mathcal{N}_i
        \right)
        &=
        \max_{\mathbf{M}}
        \mathcal{R}_{IP}^{JM}
        \left(
            \sum_i p(i)\mathcal{N}_i;\mathbf{M}
        \right) \\
        &\leq
        \max_{\mathbf{M}}
        \sum_i p(i)
        \mathcal{R}_{IP}^{JM}(\mathcal{N}_i;\mathbf{M}) \\
        &\leq
        \sum_i p(i)
        \max_{\mathbf{M}}
        \mathcal{R}_{IP}^{JM}(\mathcal{N}_i;\mathbf{M}) \\
        &=
        \sum_i p(i)
        \mathcal{R}_{IP}^{JM}(\mathcal{N}_i).
    \end{aligned}
    \]
    This concludes the proof.
    
\section{Seesaw algorithm}
    \label{appendix:seesaw}
    In this section, we introduce the seesaw algorithm of finding the potentially optimal measurement assemblage for \(\mathcal{R}^{JM}_{IP}(\mathcal{N})\).
    
    In section~\ref{subsec:quantifying}, we introduce the optimization problem in Eq.~\eqref{eq:optimization JM} and its primal and dual SDP in Eqs.~\eqref{eq:optimization JM SDP}, and ~\eqref{eq:optimization JM SDP dual}, respectively. To obtain Eq.~\eqref{eq:robustness JM}, we need to optimize the input measurement assemblage \(\mathbf{M}\).
    Thus, we arrive the following optimization problem in terms of the dual formula~\eqref{eq:optimization JM SDP dual}, \(\mathcal{R}_{IP}^{JM}(\mathcal{N})\) can be express as the following optimization problem
    \begin{equation}
        \begin{aligned}
            \text{Given } &\mathcal{N}\\
            \text{Find } &\max_{\mathbf{M}, \mathbf{Y}, Z, S} \sum_{a, x}\operatorname{Tr}[Y_{a|x}\mathcal{N}^\dagger(M_{a|x})] = 1+\mathcal{R}^{JM}_{IP}(\mathcal{N})\\
            \text{s.t. } & Y_{a|x} = Y^\dagger_{a|x}, \quad \forall a, x\\
            &M_{a|x} \geq 0, \quad M_{a|x} = M^\dagger_{a|x}, \quad \forall a, x\\
            &\sum_a M_{a|x} = \openone, \quad \forall x\\
            &\sum_{a, x} M^T_{a|x} \otimes Y_{a|x} \geq \openone \otimes S \\
            &\operatorname{Tr}[S] \geq 0, \quad S = S^\dagger\\
            &Z \geq \sum_{a, x} D(a|x, \lambda)Y_{a|x} \\
            &\operatorname{Tr}[Z] \leq 1, Z = Z^\dagger.
            \label{eq:robustness JM bilinear optimization}
        \end{aligned}
    \end{equation}
    This optimization problem contains the bilinear expression \(\operatorname{Tr}[Y_{a|x}\mathcal{N}^\dagger(M_{a|x})]\). We can decouple the constraints into those which only involve \(\mathbf{Y}\), and those that only involve \(\mathbf{M}\). Therefore, we arrive the following see-saw algorithm:
    
    First, by giving a measurement assemblage \(\mathbf{M}^0 = \{M_{a|x}^0\}_{a, x}\), we  optimize a \(\mathbf{Y}\) by the following SDP:
    \begin{equation}
        \begin{aligned}
            \text{Given } &\mathcal{N}, \mathbf{M}^0\\
            \text{Find } &\max_{\mathbf{Y}, Z, S} \sum_{a, x}\operatorname{Tr}[Y_{a|x}\mathcal{N}^\dagger(M_{a|x}^0)] - 1 \\
            &=\mathcal{R}^{JM}_{IP}(\mathcal{N}; \mathbf{M})\\
            \text{s.t. } & Y_{a|x} = Y^\dagger_{a|x}, \quad \forall a, x\\
            &\sum_{a, x} \left[M^0_{a|x}\right]^T \otimes Y_{a|x} \geq \openone \otimes S \\
            &\operatorname{Tr}[S] \geq 0, \quad S = S^\dagger\\
            &Z \geq \sum_{a, x} D(a|x, \lambda)Y_{a|x} \\
            &\operatorname{Tr}[Z] \leq 1, Z = Z^\dagger
            \label{eq:sub problem 1}.
        \end{aligned}
    \end{equation}
    We denote the optimal solution \(\mathbf{Y}\) as \(\mathbf{Y}^0\). Next, we consider \(\mathbf{Y}^0 \) as the input of the following SDP
    \begin{equation}
        \begin{aligned}
            \text{Given } &\mathcal{N}, \mathbf{Y}^0\\
            \text{Find } &\max_{\mathbf{M}, S} \sum_{a, x}\operatorname{Tr}[Y^0_{a|x}\mathcal{N}^\dagger(M_{a|x})] = 1+\mathcal{R}^{JM}_{IP}(\mathcal{N})\\
            \text{s.t. } &M_{a|x} \geq 0, \quad M_{a|x} = M^\dagger_{a|x}, \quad \forall a, x\\
            &\sum_a M_{a|x} = \openone, \quad \forall x\\
            &\sum_{a, x} M^T_{a|x} \otimes Y^0_{a|x} \geq \openone \otimes S \\
            &\operatorname{Tr}[S] \geq 0, \quad S = S^\dagger.
            \label{eq:sub problem 2}
        \end{aligned}
    \end{equation}
    We denote the optimal \(\mathbf{M}\) by \(\mathbf{M}^1 = \{M^1_{a|x}\}_{a|x}\). Thus, $\mathbf{M}^1$ can obtain a higher value in the objective function than $\mathbf{M}^0$. We can now consider $\mathbf{M}^1$ as the new input of the SDP in Eq.~\eqref{eq:sub problem 1}.
    This process is iterated until the sequence \(\{\mathbf{M}^i, \mathbf{Y}^i\}\) converges within the limits of numerical precision, which concludes our see-saw algorithm.

    %\newpage

    %\subsection{method for finding best pre-filter}
    %In this section, we will show how to find the Bob's side local filtering operation that maximize the steerability of state assemblage.
    %To achieve this, we need introduce the dual SDP formulation of incompatibility robustness
    %\begin{equation}
    %    \begin{aligned}
    %        \text{Given }& B\\
    %        \text{Find }&\max \operatorname{Tr} \left[\sum_{a, x} D(a|x, \lambda)\omega_{a|x}  \right] \eqcolon 1 + \mathcal{R}_I(B) \\
    %        \text{s.t. }& \eta \geq \sum_{a, x} D(a|x, \lambda)\omega_{a|x} \quad \forall\lambda, \\
    %        &\omega_{a|x} \geq 0, \quad \operatorname{Tr}(\eta)=1
    %    \end{aligned}
    %\end{equation}
    %where \(D(a|x, \lambda)\)  is the deterministic postprocessing, \(\mathbf{\omega} = \omega_{a|x}\) is  \(\eta\) is 

    %\begin{equation}
    %    \begin{aligned}
    %        \mathbf{\sigma}^{max} = \eta^{1/2}\rho_B^{-1/2}\mathbf{\sigma}\rho_B^{-1/2}\eta^{1/2}
    %    \end{aligned}
    %\end{equation}

\end{document}